# Towards Federated, Green, and Resilient 6G Non-Terrestrial Networks

**Sarath Babu[1], Victor Baños-Gonzalez[2], Mario Cordina[3], Debabrata Dalai[4], Tomaso de Cola[5], Franco Davoli[6], Etienne Victor Depasquale[3], Ashutosh Dutta[7], Hesham ElBakoury[8], Michael A. Enright[9], Giovanni Giambene[10], Sumit Goswami[11], Ramesh Gupta[12], Wael Jaafar[13], Eman Hammad[14], B. S. Manoj[4], Tony Li[15], Manuel M. H. Roth[5], Paresh Saxena[16], Pat Scanlan[17], Zhili Sun[18], Daniele Tarchi[19], Saviour Zammit[3]**

[1]Iowa State University, USA
[2]iABG, Germany
[3]Department of Communications & Computer Engineering, University of Malta
[4]Indian Institute of Space Science and Technology, India
[5]German Aerospace Center (DLR), Germany
[6]DITEN-University of Genoa / CNIT National Laboratory of Smart and Secure Networks (S2N), Genoa, Italy
[7]Johns Hopkins University, USA
[8]Independent Consultant, Santa Clara, CA, USA
[9]Quantum Dimension, Inc., USA
[10]Department of Information Engineering and Mathematics, University of Siena, Italy
[11]Centre for Quantum Engineering, Research and Education, TCG Crest, Kolkata 700091, India
[12]IEEE MTT-S, USA
[13]École de technologie supérieure, Montreal, Canada
[14]Texas A&M University, USA
[15]Hewlett Packard Enterprise, USA
[16]BITS Pilani, Hyderabad Campus, India
[17]Scanlanavia.com, Ireland
[18]University of Surrey, UK
[19]Department of Information Engineering, University of Florence, Italy

Corresponding author: Giovanni Giambene (e-mail: giovanni.giambene@unisi.it).

**ABSTRACT** This study focuses on future Non-Terrestrial Networks (NTN) integrated with Terrestrial Networks (TN) for future 5G/6G systems. NTN envisions a 3D architecture, where Low Earth Orbit (LEO) satellite networks will play a key role in bridging the digital divide, complementing the gradual terrestrial 5G/6G rollout concentrated in high-density and high-traffic areas, by ensuring service continuity across broad geographic regions and providing coverage in case of emergencies or in remote areas. In this context, we address networking issues for the integration and federation of Terrestrial and Non-Terrestrial Network (T-NTN) in line with the IMT-2030 vision, focusing on interoperability, spectrum coexistence, unified control and management, and service continuity. Federation is a complementary approach to integration that enables distinct satellite systems to cooperate through agreements, potentially unified satellite terminals, and common resource management. We show that system federation significantly enhances both latency performance and connectivity robustness compared with non-federated LEO architectures. This paper also investigates the challenges and possible solutions for adopting the Open-RAN architecture for T-NTN, including routing options for mega-LEO systems, edge intelligence, and energy efficiency as critical elements for sustainability. Finally, we address the security, privacy, and resilience aspects of federated T-NTN architectures with emphasis on zero-trust, secure routing, trustworthy edge intelligence, Post-Quantum Cryptography (PQC), and Quantum Key Distribution (QKD).

## I. INTRODUCTION

Non-Terrestrial Networks (NTNs) have the potential to bridge connectivity gaps in remote areas and provide resilient communication infrastructure in disaster-prone regions. These systems encompass aerial technologies operating at various altitudes and exhibit distinct coverage and propagation delay characteristics. In addition to Low-Earth Orbit (LEO), Medium-Earth Orbit (MEO), Geostationary Earth Orbit (GEO) satellites, Uncrewed Aerial Vehicles (UAVs) and High-Altitude Platforms (HAPs) are considered. The latter provide more focused coverage and represent a good solution for low-cost local/regional coverage. LEO systems are now very popular and typically deployed using mega-constellations of thousands of satellites. As an example, Figure 1 shows that the median downlink access bit rates achieved by users of the LEO Starlink system have steadily increased over the years as the constellation has expanded through the progressive deployment of satellites. As a result, the access performance of these systems is now approaching that of terrestrial mobile broadband in the US, further stimulating interest in NTN solutions capable of providing broadband connectivity in areas where terrestrial mobile networks are unavailable or economically impractical.

A key emerging paradigm in this context is Direct-to-Device (D2D), also referred to as Direct-to-Cell (D2C), which enables satellites to communicate directly with conventional smartphones and IoT devices, without requiring specialized user terminals with large antennas. By effectively acting as "cell towers in space," these satellite systems aim to eliminate coverage gaps and extend connectivity to underserved and remote areas worldwide. By connecting regular smartphones to satellites, D2C promises to eliminate dead zone coverage. Companies such as Starlink, AST SpaceMobile, Amazon LEO, OneWeb, and GlobalStar have recently pushed for D2C satellite services [2],[3]. Satellite D2D connectivity has accelerated collaboration between terrestrial mobile operators and satellite providers, with numerous partnerships and the ongoing development of device-compatible chips, followed by players such as AST SpaceMobile and Lynk Global. More recently, satellite connectivity has also started to enter the automotive sector through the first commercial integrations of NTN capabilities into production vehicles. For example, the collaboration announced between BMW and Viasat represents an important milestone toward native in-vehicle satellite connectivity.

Despite strong momentum, challenges remain in the satellite segment. To close the link budget with standard mobile handsets, satellites must support high-power amplification and employ very large, high-gain antennas or electronically steerable phased arrays capable of forming narrow, highly Effective Isotropic Radiated Power (EIRP) beams over wide coverage areas. These antennas often exceed those used in traditional broadband LEO satellites and require innovative deployable structures such as foldable origami-inspired antennas to satisfy launch constraints. In addition, the availability and cost of smartphones supporting satellite connectivity remain limited, while spectrum regulation and international coordination are becoming increasingly important to ensure the widespread deployment of NTN services. In this context, recent initiatives such as the Global System for Mobile Communications Association (GSMA) Satellite Regulatory Playbook advocate technology-neutral regulatory frameworks and regulatory parity between terrestrial and satellite operators to facilitate the seamless integration of NTN services into future communication ecosystems.

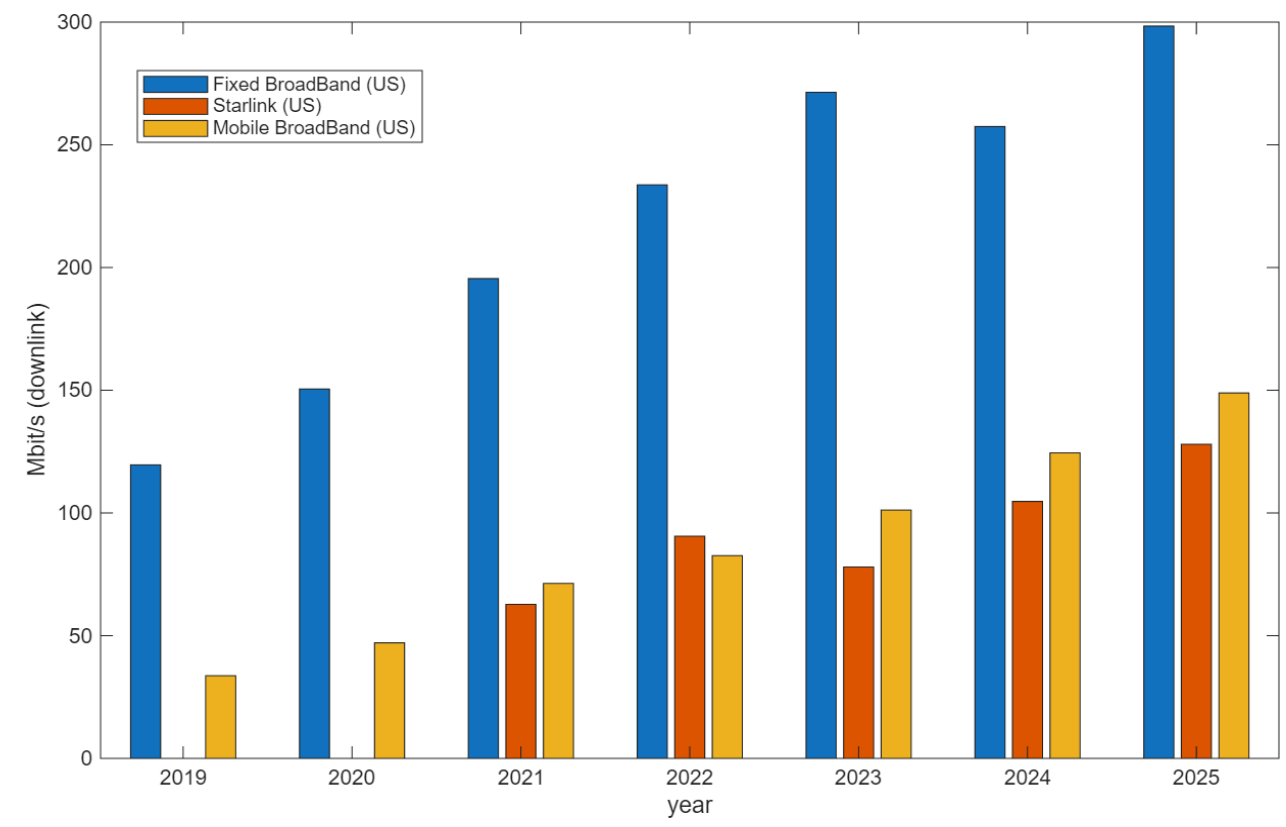

**FIGURE 1. Downlink traffic capacity (median bit rate) comparison for fixed broadband, LEO Starlink, and mobile broadband in recent years till 2025 in the US based on Ookla data [1].**

Many satellite operators complement their systems by combining assets operating in different orbital regimes either through corporate integration or cooperation agreements with other operators. Examples of this type of integration include Eutelsat's acquisition of OneWeb, which combines GEO and LEO systems, and SES's ownership of O3b, which integrates GEO and MEO capabilities. Amazon Leo recently initiated the acquisition process for the GlobalStar LEO satellite system. Simultaneously, cooperation-based approaches are emerging, such as the partnership announced between SES and Starlink, to provide seamless maritime connectivity via a Software-Defined Wide Area Network (SD-WAN). Thus, software control is used to intelligently manage, route, and optimize traffic across multiple orbit systems. The SD-WAN treats different connectivity options as a unified policy-driven network.

One of the primary advantages of multi-orbit systems is their ability to optimize the trade-offs among latency, coverage, and capacity. GEO satellites provide persistent, wide-area coverage and high aggregate capacity but suffer from a high propagation delay, which limits their suitability for latency-sensitive applications. By contrast, LEO and MEO systems offer significantly reduced latency owing to their lower orbital altitudes, enabling improved support for real-time and interactive services. SD-WAN is particularly well suited to multi-orbit satellite architectures because it can seamlessly switch between or combine orbits, mask link disruptions (e.g., LEO handovers), optimize user experience and Quality-of-Service (QoS) without manual intervention, and improve system robustness. These new multi-orbit systems enable a 3D architecture with novel capabilities but also require appropriate routing schemes.

The work presented in this paper is the result of a collaborative effort conducted within the International Network Generations Roadmap (INGR) Satellite Working Group (WG) of IEEE, whose mission is to investigate the challenges, opportunities, and enabling technologies for the integration of Terrestrial Networks (TNs) and NTNs in the context of 5G and beyond, with a particular focus on 6G systems. Building on the rapid evolution of wireless communication and the need for ubiquitous global connectivity, this WG has addressed the role of NTNs (including satellite constellations, HAPs, and UAVs) as key components of future network architectures to bridge the digital divide and support emerging vertical applications.

Over the years, the INGR Satellite WG has produced several editions of a comprehensive *roadmap report* that analyzes integration issues from multiple perspectives, reflecting both technological advances and standardization activities, notably within the 3rd Generation Partnership Project (3GPP). These roadmaps cover a broad range of topics, including federated and open network architectures, antenna and radio-frequency design, waveform and modulation schemes, routing and mobility management for mega-LEO satellite systems, and edge intelligence. Collectively, these efforts aim to outline a technology-agnostic roadmap toward integrated terrestrial-satellite networks capable of delivering enhanced capacity, resilience, and global coverage in support of future services and the United Nations' Sustainable Development Goals [4],[5].

## II. OVERVIEW OF CURRENT STATUS

Recent studies and operational evidence have highlighted the inherent fragility of large LEO mega-constellations when considered in isolation. High satellite densities lead to frequent close approaches and necessitate continuous collision-avoidance maneuvers, whereas extreme space weather events, such as severe solar storms, can disrupt satellite Command and Control (C&C) and require rapid reconfiguration. These factors further motivate federated multi-orbit architectures, where traffic can be dynamically redistributed across heterogeneous systems to enhance overall network resilience and service continuity. Table 1 provides an overview of the main LEO satellite systems for mobile communications, including salient characteristics such as orbit type, frequency bands, and the number of satellites currently in orbit.

Next-generation networks aim to provide global, low-latency services by integrating TN and NTN. Proprietary satellite networks pose several challenges, including vendor-specific ecosystems, constrained resources, and limited interoperability, which can lead to limited coverage, inefficient resource utilization, and inability to support seamless global services. Given the diversity of satellite system tenants, interoperable operations have become challenging. In essence, *interoperability* enables different technologies to work together, whereas *federation* manages interconnected networks as a single, unified system for pervasive, intelligent connectivity. Interoperability is the antithesis of closed solutions, adopting a standardized framework that enables diverse systems (developed by different vendors and operators) to interconnect and interoperate, ensuring compatibility, scalability, and long-term sustainability of the overall ecosystem. On the other hand, federation is defined as collaboration among multiple independent operators and domains (terrestrial mobile operators, satellite constellation providers, and aerial platform providers) to share resources, enable roaming, and jointly orchestrate services.

IMT-2030, this is the term adopted by ITU-R to designate 6G systems, is envisioned as a globally integrated communication framework that seamlessly unifies TNs and NTNs, including satellite systems across multiple orbits. Specifically, interoperability extends beyond basic interconnection [6] to encompass service continuity, spectrum coexistence, and unified management across TN and NTN platforms. This is particularly critical, as satellite networks transition from standalone infrastructures to integral components of the global mobile ecosystem. In addition, 3GPP has a key role in making 5G/6G a system with TN and NTN components integrated at different levels, ranging from the most basic, where they share a common core network, to the most advanced, where they share a common core and access network. NTN systems are part of the 5G/6G standardization established by 3GPP Releases 17 and 18 and closed in June 2022 and 2024, respectively, with a focus on transparent satellites, whereas regenerative satellites are addressed in Release 19 (frozen in 2026) and Release 20.

In 5G TN, federation and interoperability are defined by edge/cloud federation, multi-operator roaming, network slicing across domains, and open interfaces. For instance, the 5G Infrastructure Public Private Partnership (5GPPP) [7] emphasizes multi-domain management and orchestration. Federation and interoperability in the satellite-terrestrial domain pose several challenges, highlighting the critical role of standardization in enabling seamless integration. These challenges include delay heterogeneity (e.g., long delays for GEO and variable delays for MEO/LEO), mobility and cross-domain handovers (horizontal/vertical across terrestrial, aerial, and satellite segments), resource virtualization and management, the transition from heterogeneous to unified interface frameworks, and security.

The regulatory framework supporting the integration of TN and NTN is evolving worldwide, although with different priorities across regions. In the United States, the Federal Communications Commission (FCC) has promoted a flexible regulatory framework to accelerate commercial NTN deployment and D2C services through streamlined satellite licensing and spectrum-sharing mechanisms.

China has adopted a centrally coordinated approach, with NTN development supported through national programs such as the Guowang and SpaceSail constellations, coordinated spectrum planning, and regulatory oversight by the Ministry of Industry and Information Technology (MIIT) to strengthen the national communications infrastructure.

India is progressively improving its regulatory framework through the Indian Telecommunications Act (2023) and recent spectrum allocation policies, enabling satellite broadband and NTN services.

In Europe, the emphasis has instead been placed on regulatory harmonization, interoperability, and digital sovereignty to foster an integrated single market for satellite communications. However, regulatory fragmentation and the lack of common spectrum conditions for satellite services still hinder the development of a genuine European single market. To address these challenges, the recently approved European Digital Networks Act (DNA) establishes a harmonized legal framework for both TN and NTN, strengthening resilience and security [8]. One of the measures introduced by the EU DNA is the Union Preparedness Plan for Digital Infrastructures, which strengthens the resilience of both TN and NTN by addressing risks such as natural disasters, cyberattacks, and jamming.

Moreover, the forthcoming European GOVSATCOM Hub [9] is among the first concrete implementations of a federated satellite communication architecture, in which heterogeneous space and ground resources are managed and accessed through a unified service interface. Rather than relying on a single satellite system or operator, the GOVSATCOM Hub aggregates the capacities of multiple commercial and governmental satellite providers, potentially spanning different orbits, frequency bands, and system architectures. Through a centralized booking and service management interface, authorized government users can request, reserve, and activate communication services in accordance with predefined security, availability, and performance requirements without directly interacting with the underlying satellite operators or technologies. As such, the European GOVSATCOM Hub can be seen as a precursor to future interoperable AI-enabled network orchestration platforms envisioned for IMT-2030, in which TN and NTN communication resources are federated across organizational and technological boundaries. This demonstrates how standardized interfaces enable system integration. The 5G-HUB project plans to demonstrate the integration of TN and NTN systems to ensure service continuity within the GOVSATCOM ecosystem [10].

Efforts and funding in Europe, the USA, Asia, and other regions are driven by their respective governments and continental organizations. This scenario presents several challenges, such as interoperability across countries in terms of regulations, vendors, platforms, and different domains. The focus areas are clear: sustainability, technological autonomy of regions or countries, and open-source infrastructure oriented toward national or regional sovereignty. Flagship project trends include seamless orchestration and management across TN and NTN, service-based architectures for NTN, cloud-native architectures for NTN, edge intelligence [including both Artificial Intelligence (AI) and Machine Learning (ML) available at the network edge], sovereign-federated and green cloud and edge elastic ecosystems, NTN and computing orchestration in distributed systems, interoperability frameworks for edge networks, open standards for edge networks, and carbon-neutral and sustainable edge infrastructure. The projects in Table 2 present reference standards, architectures, and production-ready technologies that aim to address the energy-efficient, interoperable edge with NTN, testbeds, systems, and frameworks, thereby providing structure and shape to the green networking landscape for the future.

## III. SPECTRUM ALLOCATIONS

Spectrum allocation remains one of the most fundamental enablers and constraints for the implementation and expansion of 5G and 6G networks. The ITU-R World Radio Conference (WRC) process identifies the spectrum for IMT and/or designates the Fixed Satellite Systems (FSS)/Mobile Satellite Systems (MSS) bands under which NTN can operate. 3GPP then defines the air interface requirements within that spectrum, with assigned band numbers and radio specifications for TN and NTN. World Administrative Radio Conference (WARC-92), organized by ITU-R, played a pivotal role in identifying spectrum for LEO mobile satellite systems (both 'Little LEO' data/position services in VHF/UHF, and 'Big LEO' voice services in L/S-band), laying the groundwork for Iridium, Globalstar, and other similar programs [23]-[25]. An additional 230 MHz of spectrum in the 1,885–2,025 MHz and 2,110–2,200 MHz ranges was allocated for IMT-2000. This spectrum directly defines the 3GPP bands corresponding to the 2100 MHz Frequency Division Duplexing (FDD) bands used globally for Personal Communications Services (PCS). Driven by the capacity demands of successive generations of networks and the integration of satellites into the cellular standards framework, the industry and WRC-driven spectrum allocations have dramatically expanded from 230 MHz at WARC-92 (1992) to a globally harmonized spectrum of over 14.75 GHz after WRC-19 (2019), with the addition of millimeter-wave spectrum across five band ranges. In parallel with the ITU-R and WRC harmonization activities, the European Union is considering a more coordinated NTN and satellite spectrum framework under the proposed DNA [7], aiming to reduce the current fragmentation caused by national licensing approaches. Table 3 lists the current standardized 3GPP NTN bands and their corresponding conventions. These bands broadly correspond to the emerging NTN frequency-range classification, where L/S MSS bands are associated with F1, Ka-band FSS bands with F2, and the future V/Q-band allocations under investigation are expected to constitute F3. In addition to these standardized bands, 3GPP Release-19 has approved a work item addressing NTN operation in Ku-band FSS/BSS (~12.75–14.5 GHz for Uplink, UL, and ~10.7–12.75 GHz for Downlink, DL); however, the corresponding band numbers have not yet been finalized. NTN operation in the V/Q-band FSS ranges (47.2–51.4 GHz UL and 37.5–42.5 GHz DL) is currently under study for IMT-2030 and WRC-27.

TABLE 1: COMPARISON OF MAJOR COMMERCIAL LEO SATELLITE SYSTEMS AS OF JULY 2026.

| System / Constellation | Operator | Deployment Status | Frequency Bands Used | Service Status & Notes |
|---|---|---|---|---|
| Starlink | SpaceX | >12000 operational LEO satellites; currently using 8 shells with different orbital parameters | Ku-band, Ka-band; exploring V/E-band (mmWave) | Global broadband service; D2D messaging with progressive expansion towards voice and data services |
| Amazon Leo (Project Kuiper) + | Amazon | 370 operational LEO satellites - Deployment in progress (multi-shell planned) | Ka-band, V-band for future expansions | Broadband-focused LEO constellation; service rollout initiated |
| OneWeb LEO + Eutelsat GEO | Eutelsat / OneWeb | 656 operational LEO satellites + GEO satellites | Ku-band, Ka-band | Enterprise, government, and mobility services; multi-orbit integration |
| AST SpaceMobile | AST SpaceMobile | 9 operational LEO satellites – gradual deployment | L-band / S-band / terrestrial cellular bands (e.g., ~850 MHz) | D2C connectivity for unmodified, standard smartphones |
| Lynk Global | Lynk Global | Early deployment of LEO satellites/trials | Terrestrial mobile spectrum (sub-3 GHz cellular bands) | Direct messaging and limited data for unmodified, standard smartphones |
| Telesat Lightspeed | Telesat | No operational LEO satellite yet - planned launches starting 2026–2027 | Ka-band | Planned global broadband LEO constellation; enterprise and government focused |
| SES (O3b + GEO) | SES | ~10 Operational O3b mPower MEO + SES GEO fleet / multi-orbit system | Ka-band (MEO), C/Ku/Ka (GEO) | Mature GEO+MEO architecture with differentiated latency and capacity |
| China SatNet "Guowang" (GW) | China SatNet | ~170 operational LEO satellites (multi-shell constellation with planned >13000 satellites) | Ku-band, Ka-band | Rapidly expanding |

TABLE 2: COMPARISON OF PROJECTS ON NTN SYSTEMS.

| Project Name | Trend | Objective and Results | Relation to Green NTN Networking |
|---|---|---|---|
| NTT Data (US, EU, and Global) Edge AI platform [11] | Edge platform, sovereign infrastructure with NTN | **Objective**: Provide an edge AI platform enabling real-time AI processing in smart infrastructure, industry, and smart cities<br>**Results**: Early deployments in the EU and the US with a focus on maintenance, GenAI, and security. | Supports scalable open, neutral edge AI, with interoperability across telcos, 5G, and IoT environments. |
| European Union initiatives: SmartEdge [12], 6Green [13], 6G-NTN [14], 5G-STARDUST [15], Ether [16], NexaSphere [17], Unity-6G [18], 5G-HUB [9] | Adaptive edge intelligence and swarm IoT, AI-powered, energy-efficient, interoperable edge-cloud systems, green network, sustainable 6G, TN-NTN integration and NTN federation | **Objective**: Efficient, intelligent edge computing with swarm intelligence for IoT, industry, cities, and healthcare. Creation of a service-based, holistic ecosystem that enables sustainability. Develop 6G non-terrestrial and satellite, air, and ground networking with focus on seamless integration, green design principles, and energy efficiency.<br>**Results**: A market-ready low-code toolchain for edge AI, secure, distributed authentication for IoT, and NTN. Demonstrates end-to-end green elasticity for NTN vertical applications and a dynamic adaptation workload across the edge-cloud continuum. Improved technologies, protocols, and architectures for future 6G and NTN, including edge-offloading and AI-powered edge intelligence for combined terrestrial and non-terrestrial networks. | Green real-time IoT and NTN use cases through intelligent, cooperative edge nodes. Integration in edge-cloud, AI, sustainability for NTN, TN-NTN integration |
| European Space Agency (ESA) initiatives: NTN edge, EDGECOLB, GreenNet, SatNetGreen, GreenComm, NTN-CO, NTN-CF, GreenSatNet [19] | NTN and satellite green communications, cloud-edge computing | **Objective**: Develop an NTN system with edge computing capabilities, develop an eco-friendly satellite communication network, investigate green communication technologies for future space missions, and develop an NTN network with cognitive capabilities and cloud-fog computing capabilities.<br>**Results**: Improved latency, capacity, and enhanced security, reduced carbon footprint, reduced energy consumption, and improved adaptability. | Green satellite and NTN: sustainable communication networks, utilizing cognitive edge-fog-cloud computing capabilities for NTN use cases. |
| China National Space Administration (CNSA) Initiatives: BeiDOU Navigation Systems in NTN Context [20], China-Europe SMILE (ESA-CNSA) [21] | NTN, Open-RAN, edge | **Objective**: 5G NTN integration through Open RAN for challenging environments, supporting centimeter-level outdoor accuracy. SMILE extends the NTN concept into scientific NTNs, demonstrating advanced O-RAN for NTN orchestration through the space-time project<br>**Results**: Field test reported 2 cm horizontal deviation outdoors and 0.11 m average error indoors, demonstrating the feasibility of NTN-augmented positioning accuracy. | NTN orchestration with Open RAN for outdoor accuracy |
| JAXA (Japan Aerospace Exploration Agency) [22]<br>TANSAX (Space exploration Innovation Hub Center)<br>LUCAS (Optical Inter-satellite Communication) | Satellite communications and NTN | **Objective**: Supports infrastructure for Moon and Mars communication data relay. Combines space mission goals with commercial terrestrial applications. Enable communications between LEO and GEO satellites.<br>**Results**: Precursor to future ISL network and regenerative payload in 3GPP NTN phase 3 (Release +19) for real-time telemetry and high-data applications. | Deploys advanced communication arrays for NTN and extraterrestrial links, with a focus on fulfilling high-performance telemetry requirements |

TABLE 3: NTN BANDS WITH NUMBERING CONVENTION

| 3GPP Band | UL (GHz) | DL (GHz) | Spectrum | Rel. | Notes |
|---|---|---|---|---|---|
| n255 | 1.626 – 1.661 | 1.525 – 1.559 | L-band MSS | 17 | L-band satellite; global MSS |
| n256 | 1.980 – 2.010 | 2.170 – 2.200 | S-band MSS | 17 | S-band satellite; WARC-92 satellite component |
| n77/n78 | 5.925–6.425 | 3.7 – 4.2 | C-Band | 15 | FSS downlinks; region-dependent |
| n510 | 27.5 – 28.35 | 17.7 – 20.2 | Ka-band FSS | 18 | Ka sub-range (lower UL) |
| n511 | 28.35 – 30.0 | 17.7 – 20.2 | Ka-band FSS | 18 | Ka sub-range (upper UL) |
| n512 | 27.5 – 30.0 | 17.7 – 20.2 | Ka-band FSS | 18 | Full Ka FSS pairing — broadband VSAT / NTN |

TABLE 4: US SPECTRUM BANDS (600 MHZ – 2 GHZ) DESIGNATED FOR SUPPLEMENTAL COVERAGE FROM SPACE (SCS)

| Band | Frequency Range (Uplink/Downlink) | Band Type | Notes |
|---|---|---|---|
| 600 MHz | 663–698 MHz / 617–652 MHz | FDD | T-Mobile primary licensee; key rural coverage band |
| 700 MHz | 776–787 MHz / 746–757 MHz (A/B/C Blocks) | FDD | Wide area coverage; AT&T and others |
| 800 MHz Cellular | 824–849 MHz / 869–894 MHz | FDD | Legacy cellular band; widely deployed |
| PCS G Block | 1910–1930 MHz / 1990–2010 MHz | FDD | Key SpaceX/T-Mobile SCS band; 1910–1915/1990–1995 MHz licensed |
| AWS-H Block | 1915–1920 MHz / 1995–2000 MHz | FDD | Designated for SCS; adjacent to PCS G block |
| Broadband PCS | 1850–1910 MHz / 1930–1990 MHz | FDD | Multiple blocks; subject to SCS eligibility criteria |

### A. MOBILE SATELLITE SERVICE (MSS) D2D SPECTRUM

D2D offers a cost-effective way to extend broadband and voice connectivity to underserved and unserved communities globally. There are two fundamentally distinct spectrum strategies for delivering D2D services, each with different regulatory, technical, and commercial considerations. The first approach, denoted by MSS D2D, uses the part of the 3GPP FR1 spectrum, which is already allocated to the MSS in the ITU Radio Regulations, primarily the L band (1–2 GHz) and S band (2–4 GHz), with the possibility of being extended up to the C-Band (4-6 GHz). The second approach, referred to as IMT D2D or Supplemental Coverage from Space (SCS), enables satellites to use spectra allocated and licensed to a terrestrial mobile service (IMT). The US FCC adopted the SCS regulatory framework in March 2024 [26]. It enables satellites to communicate directly with unmodified, standard smartphones via Mobile Network Operator (MNO)-licensed cellular bands (e.g., 700 MHz and 800 MHz for Advanced Wireless Services, AWS, and Personal Communications Service, PCS) by using the existing terrestrial mobile spectrum. This makes it possible to extend mobile coverage to remote and underserved areas, including rural regions, national parks, and other locations, where terrestrial networks are neither economically viable nor technically feasible. Table 4 lists the authorized 600 MHz to 2 GHz SCS uplink and downlink bands. In the S-band (2 – 4 GHz), Mobile/IMT, MSS allocations exist in some sub-bands between 2500 and 2690 MHz. In addition, the 2483.5–2500 MHz band is globally allocated for space-to-Earth (downlink) MSS operations, and GlobalStar holds the spectrum rights for this band. Globalstar operates at 1610–1618.25 MHz (uplink, L-band) and 2483.5–2500 MHz (downlink, S-band) with its LEO constellation, supporting commercial D2D partnerships, such as with Apple.

WRC-23 did not create new spectrum allocations for D2D, but set the stage for WRC-27. The 1695–1710, 2010–2025, 3300–3315, and 3385–3400 MHz frequency bands continue to be used by the services to which they are primarily allocated. WRC-23, in accordance with resolution 252, directed WRC-27 to consider new allocations to non-geostationary MSS systems and regulatory actions in the frequency bands 1427–1432 MHz (space-to-Earth), 1645.5–1646.5 MHz (space-to-Earth/Earth-to-space), 1880–1920 MHz (space-to-Earth/Earth-to-space), and 2010–2025 MHz (space-to-Earth/Earth-to-space). WRC-27 has the important task of deciding on allocations for MSS D2D (L/S-band MSS spectrum, deployable today without new regulations) and for the use of IMT D2D (satellites' use of terrestrially licensed mobile bands) globally, based on WRC-27 allocations.

## IV. NTN SYSTEM ARCHITECTURES AND BUILDING BLOCKS

The general vision for forthcoming 6G-NTN networks is a multi-layer, multi-orbit 3D space system that is natively part of the 6G ecosystem, providing resilient, ubiquitous, and continuous coverage for society and industry. From this standpoint, the general architecture comprises a multi-folded interconnection between ground and space assets (i.e., UAVs, HAPs, and satellites), with the latter distributed across different altitudes. The general vision follows the communication paradigm illustrated in Figure 2, elaborated by the EU-funded 6G-NTN project [27], and further expanded in the NexaSphere project [28].

The expectation for unified 6G TN and NTN networks is that space networks will build on smart nodes that implement advanced processing capabilities (e.g., network functions and dedicated applications) and AI-based processors to enable the deployment of a full 5G/6G RAN and an edge concept in space.

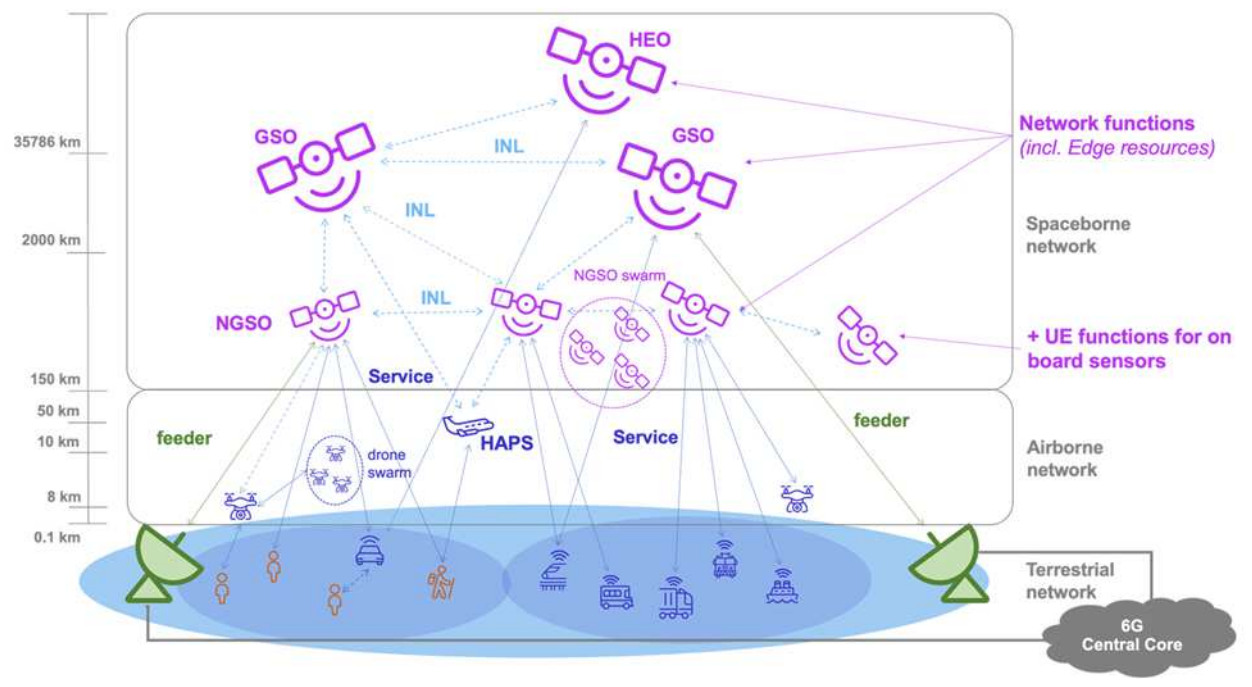

FIGURE 2. **NTN multi-layer architecture.**

To achieve this, a key challenge is the proper deployment of gNB nodes on the ground and in space, constrained by the mass, power, and computational limitations of space nodes as well as by service requirements for throughput and delay.

On the one hand, deploying monolithic gNBs in space is certainly attractive, as it would enable a natural extension of terrestrial RANs into space, but it might be challenging due to the substantial power and computational resources required to perform all signal and data processing functions in both the user and control planes.

On the other hand, the RAN disaggregation approach, which distributes gNB functionalities according to the well-known Open-RAN (O-RAN)-based architecture (depicted in Figure 3), may be more scalable from the perspective of a satellite node, but still introduces some operational challenges, as briefly elaborated in the following. Specifically, this approach is based on separating gNB functions (i.e., Radio Unit, RU, Distributed Unit, DU, and Centralized Unit, CU) into independent SW/HW components, the interworking of which is guaranteed by the service interfaces documented in the NG-RAN and 5G New Radio (NR) specifications from 3GPP. The actual application of all envisioned functional splits is not straightforward for NTN systems due to the non-negligible latency of space links, which may not meet the service interface requirements of the RU, DU, CU, and the processing functions within them. Recently, the scientific community and the space industry have focused on functional splits 1 (full gNB in space), 2 (CU/DU splitting), and 7 (RU, CU, and DU distributed).

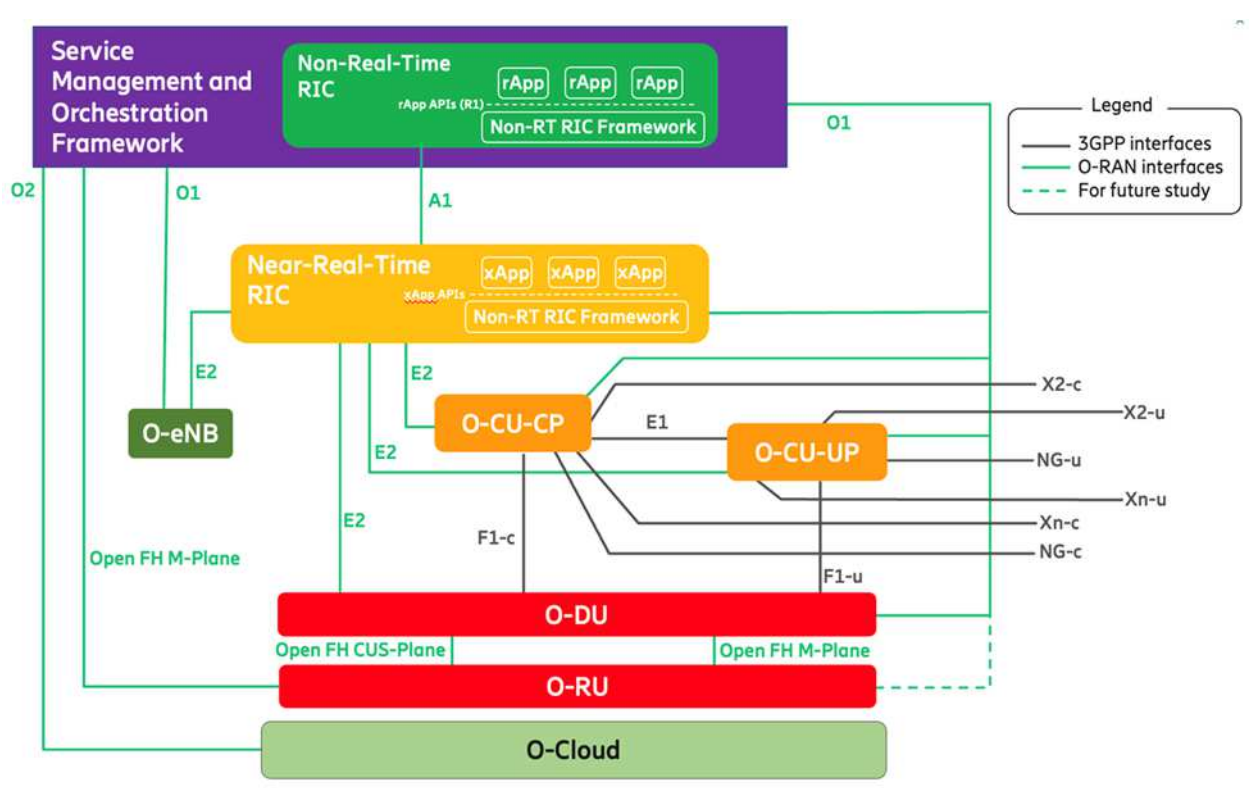

FIGURE 3. O-RAN architecture (courtesy of O-RAN alliance) [29]

The first configuration (i.e., split 1) is illustrated in Figure 4, where a fully regenerative NTN node that implements all 5G functionalities is assumed. As such, the gNB in space is responsible for fully terminating the 5G NR protocol stack, including Radio Link Control (RLC), Radio Resource Control (RRC), Packet Data Convergence Protocol (PDCP), and Service Data Adaptation Protocol (SDAP) protocol functionalities pertaining to both the user and control planes. As such, the space segment is directly involved in handover procedures, and support for switching in space can be enabled via the Xn interface, allowing interworking between gNBs onboard adjacent NTN nodes [i.e., directly interconnected through Inter-Satellite Links (ISLs)]. Moreover, IP routing in space can be enabled by implementing User Plane Function (UPF) functionalities onboard NTN nodes, thereby enabling higher-layer protocol stack implementation and paving the way toward a more service-oriented implementation of space nodes, potentially supporting additional functionalities such as edge computing in space.

The second configuration (split 2) sketched in Figure 5 presents the traditional CU/DU split, with the DU implementing all functionalities up to the RLC layer, whereas the CU is allocated PDCP-layer functionalities.

Consequently, a DU is allocated to space, whereas a CU can be deployed either in space or on the ground. If a CU is also allocated in space, it enables the deployment of an in-space distributed gNB. If a CU is implemented on the ground, the CU-DU interface will occur directly on the space-to-ground links. The main requirement for such a functional split is related to a persistent connection between the CU and DU blocks, which might be trickier when a CU is deployed on the ground because of a possible feeder link switchover; that is, when satellite nodes are no longer in visibility to the local satellite ground station and therefore must be switched over to another one, hence causing migration of the feeder link. In general, split 2 makes space routing more challenging because the user plane is split between the DU and CU, implying traffic from space to the ground and possibly back to space. A potential countermeasure to reduce ground-to-space transport traffic is to integrate user-plane functionality in space while keeping the control plane separate between the ground and space. Given that user-plane functions are much less energy-intensive than their control-plane counterparts, such an approach is considered feasible and sustainable for real NTN systems, as discussed in the 6G-NTN project.

Another possible variant in the context of a DU/CU split (i.e., split 2) is the implementation of CU functionalities in space, such that a gNB is in fact implemented in space, but through a distributed approach, that is, DU and CU are in distinct NTN nodes. In this configuration, the satellite implementing the DU functionalities establishes the service link toward the User Terminals (UEs) on the ground, whereas the DU establishes the feeder link to close the end-to-end path between the UEs and the core network running behind the satellite ground stations. CU and DU entities are interconnected across the ISLs established between adjacent satellites by implementing the NG-RAN F1 interface in space.

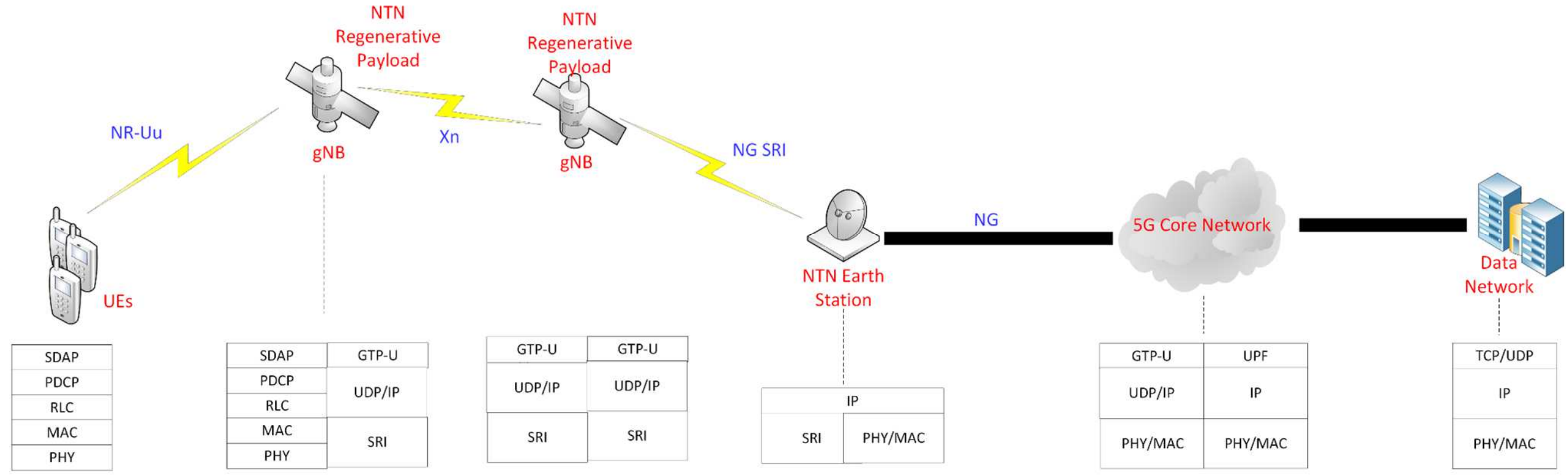

FIGURE 4. Full-gNB in space (split 1): user-plane protocol architecture and support of routing in space

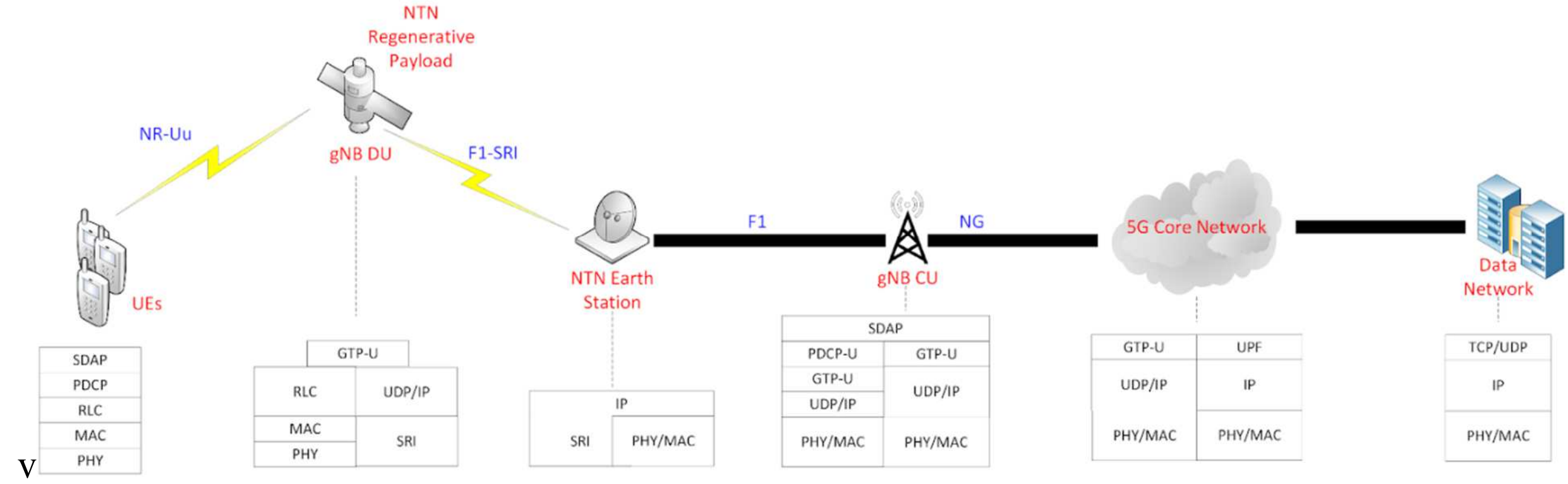

V

**FIGURE 5. CU-DU split (split 2): user-plane protocol architecture**

Although particularly appealing, this configuration requires proper engineering of the CU and DU functions and the implementation of routing schemes to ensure that the additional signaling exchanged between the CU and DU in space can be accommodated within the power and computational budget defined at the system design level.

Finally, the last attractive configuration is Split 7 (with sub-options 7.1, 7.2, 7.2x, and 7.3), which allows the maintenance of RU in space with the main physical-layer functionalities, whereas CU and DU are moved either directly on the ground or again into space, according to the in-space distributed gNB concept described previously. The main challenge of such a configuration is that communication between the RU and DU requires tight synchronization, which is particularly difficult to achieve in space when using enhanced Common Public Radio Interfaces (eCPRI). Furthermore, the exchange of control information is particularly significant because the timing requirements ultimately imply substantial bandwidth requirements on either the feeder or inter-satellite links. In particular, the availability of a large bandwidth on feeder links for control signaling could be a showstopper for systems that implement feeder links with a limited capacity.

Another important element of the O-RAN-based architecture is the definition and positioning of the RAN Intelligent Controller (RIC) and its elements, that is, near- and non-real-time RICs (sketched in the central part of Figure 3). Its implementation depends on the specific functional split adopted, although the design space is further narrowed by the fact that the near-real-time RIC operates on top of the CU block, while the non-real-time RIC typically interacts directly with the network orchestration functions in the 5G/6G core network. The overall implementation of RIC components in the NTN ecosystem is currently the object of further investigation [30],[31], especially in relation to the possible positioning of the near-real-time RIC in space and its non-real-time counterpart on the ground to facilitate easier interworking with the Service Management Orchestration (SMO) frameworks used by terrestrial and non-terrestrial operators. Furthermore, the potential of a full O-RAN-enabled architecture in space [32] is discussed, highlighting the possible evolution of NTN systems toward a more modular and flexible implementation in 6G and beyond.

## V. INTEROPERABILITY AND FEDERATION OF NTN SYSTEMS

A single constellation provides coverage and routing but faces capacity, handover, and geographic limitations. When two independent constellations cooperate (federation) and share routing and onboard computational resources, we expect potential latency improvements through load balancing, alternative routing, and reduced queuing. Furthermore, satellite constellations and terrestrial networks share spectrum and handover paths between the domains. For example, a User Equipment (UE) moving out of terrestrial coverage can perform a vertical handover to the LEO system to maintain seamless coverage. In performing this study, we assume that satellite terminals, ground gateways, and RAN nodes use open interfaces that enable multi-vendor interoperability across domains (satellite vs. terrestrial).

### *A. SYSTEM MODEL FEDERATION IN NTN SYSTEMS*

We examine federation (resource sharing) and routing strategies for satellite-terrestrial integrated 5G/6G systems, focusing on end-to-end user latency. We compare two scenarios: (*i*) A single satellite constellation serving the user; (*ii*) Two constellations that cooperate by sharing LEO satellites for routing. In the remainder of this section, we present a latency model that decomposes the end-to-end delay into access, propagation, transmission, and inter-satellite switching components, compare the latency performance with and without federation, and finally discuss the challenges and open research issues associated with realizing federated satellite networks.

We developed an analytical model to compare end-to-end latency in non-federated and federated satellite networks. We focus exclusively on propagation and transmission delays, ignoring queuing and processing overheads. The purpose is to capture situations where (*i*) the absence of a satellite or link forces a longer path in non-federated routing and (*ii*) federation enables access to additional satellites or inter-operator links that may offer shorter alternative paths.

Consider a graph G = (V, E), where V represents the set of satellite and terrestrial nodes and E represents the set of links between all nodes in the network. Each edge e ∈ E has a weight $d_e$, which consists of propagation and transmission delays, that is $d_e = \tau_e + \frac{L_e}{C_e}$, where $\tau_e$ is the propagation delay, $L_e$ is the packet size, and $C_e$ is the link capacity. Furthermore, the satellite edge availability indicator $a_e(t) \in \{0,1\}$, represents whether edge $e$ is operational at time $t$. This captures the time-varying satellite visibility and link outages. We assume two operators A and B, and use $l(e) \in \{A, B\}$. Let $h_e$ be a term used to capture the authentication, gateway translation, or routing policy overhead because of federation: when a federation is active, edges belonging to partner operators may be used for routing but may incur additional latency denoted with $h_e \geq 0$. For simplicity, we assume $h_e$ is negligible in this study. Let S be the source UE and D be the destination UE.

In the case of non-federated routing (only the edges that belong to Operator A are considered), the feasible path set $P$ at time $t$ is:

$$P_{nf}(S, D, t) = \{P \text{ from S to D}: a_e(t) = 1 \; with \; l(e) = A, \forall e \in P\}.$$

Moreover, the feasible path set $P$ at time $t$ for federated routing is:

$$P_f(S, D, t) = \{P \text{ from S to D}: a_e(t) = 1 \; with \; l(e) \in \{A, B\}, \forall e \in P\}.$$

In the non-federated case, the latency model for a given path $P$ is as follows:

$$T_{nf}(P) = \sum_{e \in P} d_e.$$

At a given time, the minimum end-to-end latency is obtained by selecting the path with the lowest latency among all feasible paths. This path is defined as follows:

$$T^*_{nf}(t) = \min_{P \in P_{nf}(S,D,t)} \sum_{e \in P} d_e .$$

Similarly, for the federation case, the latency model for a given path $P$ is

$$T_f(P) = \sum_{e \in P} w_e,$$

where $w_e = d_e + h_e$. Further, minimum federated latency is given by the equation,

$$T^*_f(t) = \min_{P \in P_f(S,D,t)} \sum_{e \in P} w_e.$$

In a non-federated system, satellite mobility may render some links unavailable ($a_e(t) = 0$). Consequently, the routing algorithm must select an alternative feasible path, which generally has a larger hop count or a longer propagation distance, thereby increasing the end-to-end latency. This increases $T^*_{nf}(t)$.

Even if an overhead $h_e$ exists in a federated system, federation may reduce the end-to-end latency by shorter feasible routing paths or reducing the hop count. The latency improvement depends on network topology, satellite availability, and the associated inter-operator overheads. The effectiveness of federation is evaluated quantitatively through the simulation results presented in Section V-C. To quantify the impact of federation, we define the latency difference as:

$$\Delta T^*(t) = T^*_{nf}(t) - T^*_f(t).$$

If $\Delta T^*(t) > 0$: Federation reduces latency, and if $\Delta T^*(t) \leq 0$: Federation increases latency.

### B. SIMULATION SETUP FOR FEDERATION IN NTN SYSTEMS

The simulation framework was implemented using an orbital-network modeling approach. Table 5 lists all network parameters considered in the simulation. Satellite orbital propagation and visibility analyses were performed using the Python library Skyfield to compute real-time satellite positions from the TLE data and the Systems Tool Kit (STK) to validate constellation visualizations. In addition, network topology construction and path computation were implemented using Python libraries: NetworkX for dynamic graph generation and SimPy for discrete-event call generation and network state evolution. Three distinct architectures were evaluated in the proposed study: (*i*) Non-Federated Iridium Next Satellite architecture; (*ii*) Non-Federated Starlink architecture; (*iii*) Federated Iridium-Starlink architecture. Although the majority of the Starlink constellation operates in inclined LEO orbital shells, Starlink also includes dedicated near-polar orbital shells to extend coverage to high-latitude and polar regions. The significantly larger size of the overall Starlink constellation compared to Iridium NEXT could bias the federation analysis. Therefore, to enable a fair and meaningful comparison, we consider only a subset of Starlink satellites belonging to its near-polar orbital shell. Since both this shell and the Iridium NEXT constellation employ near-polar orbits, the corresponding satellites exhibit similar orbital dynamics and global coverage characteristics. Restricting the analysis to a subset of Starlink satellites provides a more balanced comparison by reducing the influence of differences in constellation size while preserving comparable orbital geometry and satellite distribution.

#### 1) Non-Federated Iridium Next Satellite Architecture

In this architecture, only satellites in the Iridium Next constellation are permitted to establish ISLs. Routing paths are constrained entirely within the Iridium network. No inter-constellation communication is permitted.

**2) Non-Federated Starlink Architecture**

Similarly, in the non-federated Starlink constellation architecture, routing is restricted exclusively to the Starlink constellation. ISLs exist only between the Starlink satellites. No connectivity to the Iridium Next satellite is allowed.

**3) Federated Iridium-Starlink Architecture**

In a federated architecture, satellites from both constellations coexist within a unified routing scheme. In addition to intra-constellation ISLs, inter-constellation ISLs are enabled when satellites from different constellations are within the communication range.

TABLE 5: NETWORK PARAMETERS.

| Parameter | Value |
|---|---|
| Simulation duration | 6000 s (One Orbital Period) |
| Call arrival rate | 1/30 calls/s per UE |
| Call setup delay | 10 ms |
| Signaling delay | 10 ms |
| Earth radius | 6371 km |
| ISL bandwidth | 100 Mbps |
| Gateway-Satellite bandwidth | 50 Mbps |
| Iridium satellites | 75 |
| Starlink satellites | 121 |
| Ground stations | 10 |
| UEs per ground station | 100 |
| Total UEs | 1000 |
| Satellite time step duration | 1 s |

Figures 6 and 7 show 75 Iridium satellites with 121 polar Starlink satellites distributed in separate orbital shells. The terrestrial segment consists of globally distributed ground stations located in Australia, Brazil, China, Egypt, Germany, India, Mexico, Nigeria, the United Kingdom, and the USA. Each ground station serves 100 UEs, forming a globally distributed access network. Ground-station determination and UE deployment were considered in [33]. Two source UE–destination UE selection strategies have been considered to study Dijkstra routing performance under both average-case and worst-case conditions. These strategies are described as follows.

**1) Shortest Path – Random UE**

In the Shortest Path – Random UE (SP-RU), the source and destination UEs are selected from the active UE population using a uniform random distribution. The routing path is computed between the source and destination in the current network topology using Dijkstra's shortest-path algorithm, with propagation delay as the weight metric. This strategy represents statistically distributed traffic conditions and evaluates average-case network behavior. Figure 6 shows a routing instance under the SP-RU (red line). The path traverses a limited number of satellites and utilizes a cross-constellation route to reduce path length.

**2) Shortest Path – Farthest UE**

In the Shortest Path – Farthest UE (SP-FU), the UE pair with the maximum geographical separation is selected. The routing path is computed using Dijkstra's shortest-path algorithm. This strategy represents the worst-case spatial-routing condition. Figure 7 illustrates a routing instance under SP-FU (red line) in which the UE pair is geographically farthest apart. This path spans multiple orbital planes and crosses a large geographical arc.

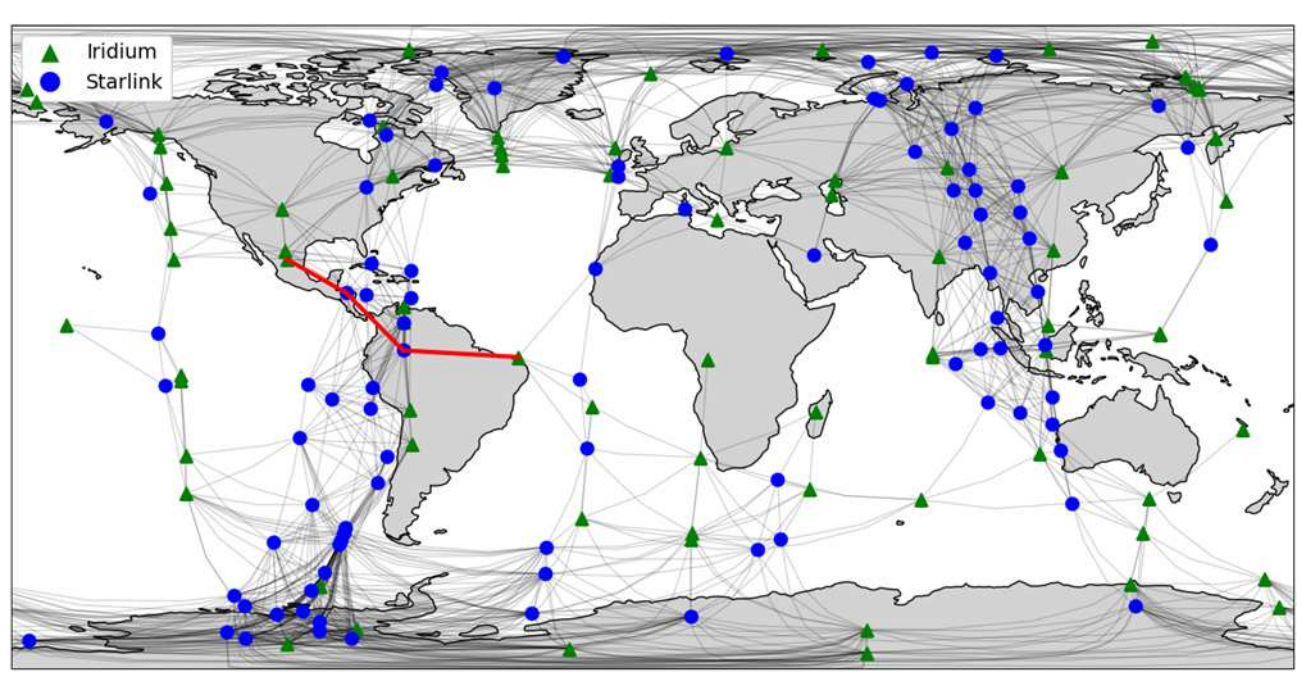

**FIGURE 6.** **Example of Shortest Path – Random UE routing. Blue circles represent the Starlink satellites, whereas green triangles represent the Iridium satellites. Here, source and destination UEs are selected from the active UE population using a uniform random distribution.**

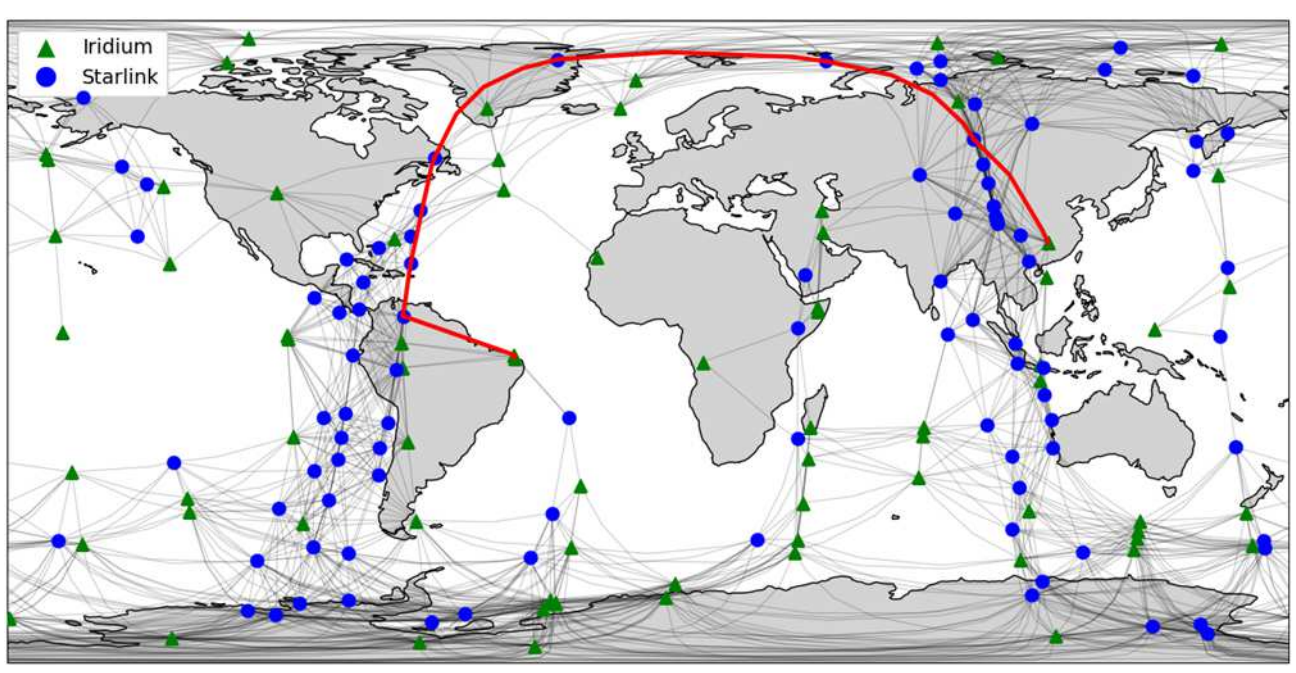

**FIGURE 7.** **Example of Shortest Path – Farthest UE routing. Blue circles represent the Starlink satellites, whereas green triangles represent the Iridium satellites. Here, the source and destination UE pair with maximum geographical separation is selected.**

### C. PERFORMANCE ANALYSIS FOR FEDERATION IN NTN SYSTEMS

This section presents a comparative evaluation of the three LEO architectures: the non-federated Iridium Next constellation, the non-federated Starlink constellation, and the federated Iridium-Starlink constellation. Each architecture was analyzed in two conditions: the Shortest Path with Random UE selection (SP-RU) and the Shortest Path with Farthest UE selection (SP-FU).

In this simulation study, the satellite network topology is updated at 1-second intervals, and all routing decisions are performed on a static topology snapshot during this period. This snapshot-based modeling approach is widely adopted in the satellite networking literature for evaluating routing and latency performance. Since the observed end-to-end latencies are significantly smaller than one second, inter-constellation links are assumed to remain valid throughout a transmission. Although finer temporal resolutions may capture additional handover events, they would affect both the federated and non-federated architectures similarly and are therefore beyond the scope of this comparative study.

Figure 8 illustrates the End-to-End (E2E) delay for the three evaluated architectures under the SP-RU and SP-FU conditions. The boxplots provide a statistical summary of the observed delays, including the median, interquartile range, whiskers, and the overall spread of the delay distribution. In the SP-RU case, the federated Iridium–Starlink architecture provides the lowest median E2E delay (68.44 ms) and the narrowest interquartile range among the three architectures. Further, the non-federated Starlink architecture achieves a lower delay of 73.44 ms than the non-federated Iridium Next architecture (88.94 ms). Furthermore, the federated architecture reduces the average E2E delay by approximately 23% compared to the non-federated Iridium Next architecture and by 6.8% compared to the non-federated Starlink architecture. This improvement is primarily attributed to the availability of inter-constellation routing paths, which increase routing flexibility and reduce the overall path length. In the SP-FU case, the entire delay distribution shifts towards higher values across all architectures because communication occurs between the geographically farthest user pairs, thereby increasing the hop count. In this case, the federated architecture continues to maintain both the lowest median and the smallest spread of delay values. Compared with non-federated architectures, the federated network exhibits a more compact delay distribution with fewer high-delay instances, indicating improved routing stability. These observations confirm that federation effectively mitigates the delay penalties associated with worst-case routing scenarios.

Figure 9 reports two complementary hop count metrics: (*i*) the number of ISL hops, which only accounts for links between satellites, and (*ii*) the E2E hop count, which includes the complete routing path from the source UE to the destination, including terrestrial access, gateways, and satellite segments. Similar to the E2E delay analysis, the boxplots reveal not only the central tendency but also the variability of the routing paths across different architectures. For the E2E routing paths, the non-federated Iridium Next architecture exhibits the highest median hop count in both cases (i.e., SP-RU and SP-FU), followed by the non-federated Starlink architecture, whereas the federated architecture consistently achieves the lowest median hop count. Further, a clear correlation can be observed between Figures 8 and 9. Architectures with lower hop counts for E2E consistently achieve lower E2E delays. Under SP-RU, the average hop count decreases from 5.29 in the non-federated Iridium Next architecture to 4.15 in the non-federated Starlink architecture and further to 3.79 in the federated architecture. Under SP-FU, the average hop counts increase to 7.51, 6.28, and 5.97, respectively. Beyond the reduction in the average hop count, the boxplots show that the federated architecture also exhibits a narrower interquartile range and shorter whiskers, which indicate the routing paths are more consistent and less sensitive to changes in user locations. In contrast, the wider distributions observed in the non-federated Iridium Next architecture suggest greater routing variability due to the limited availability of feasible shortest paths within a single constellation.

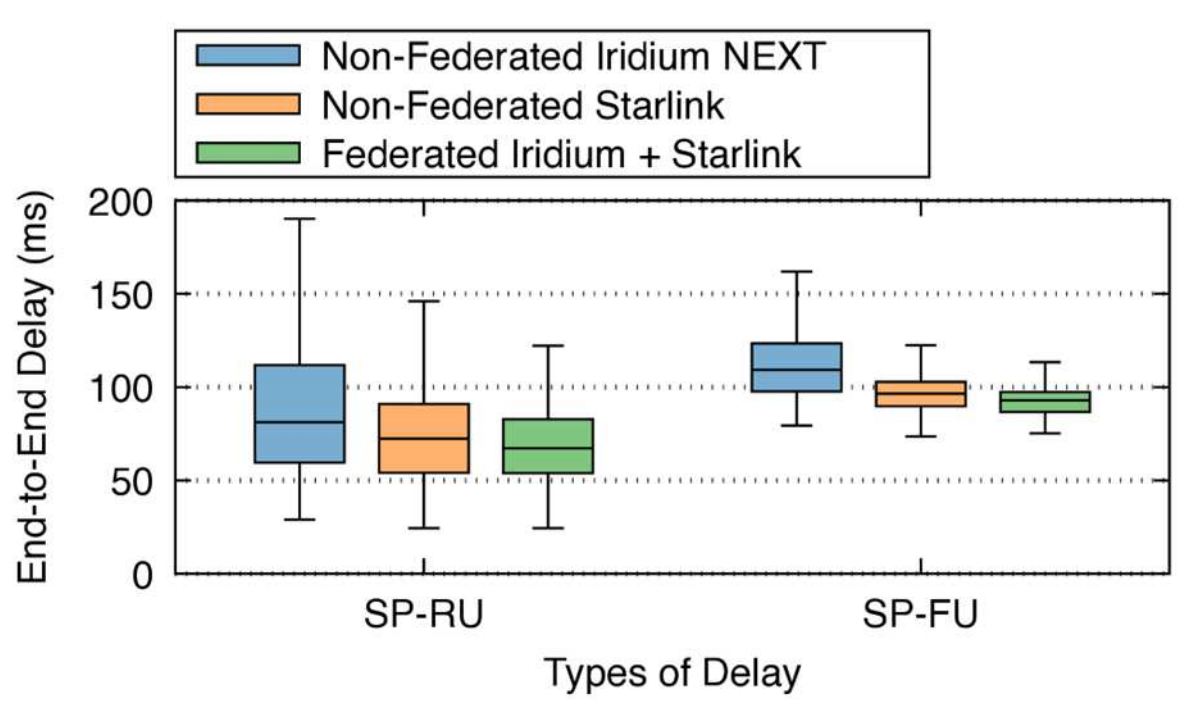

**FIGURE 8.** **Average end-to-end delay for all architectures.**

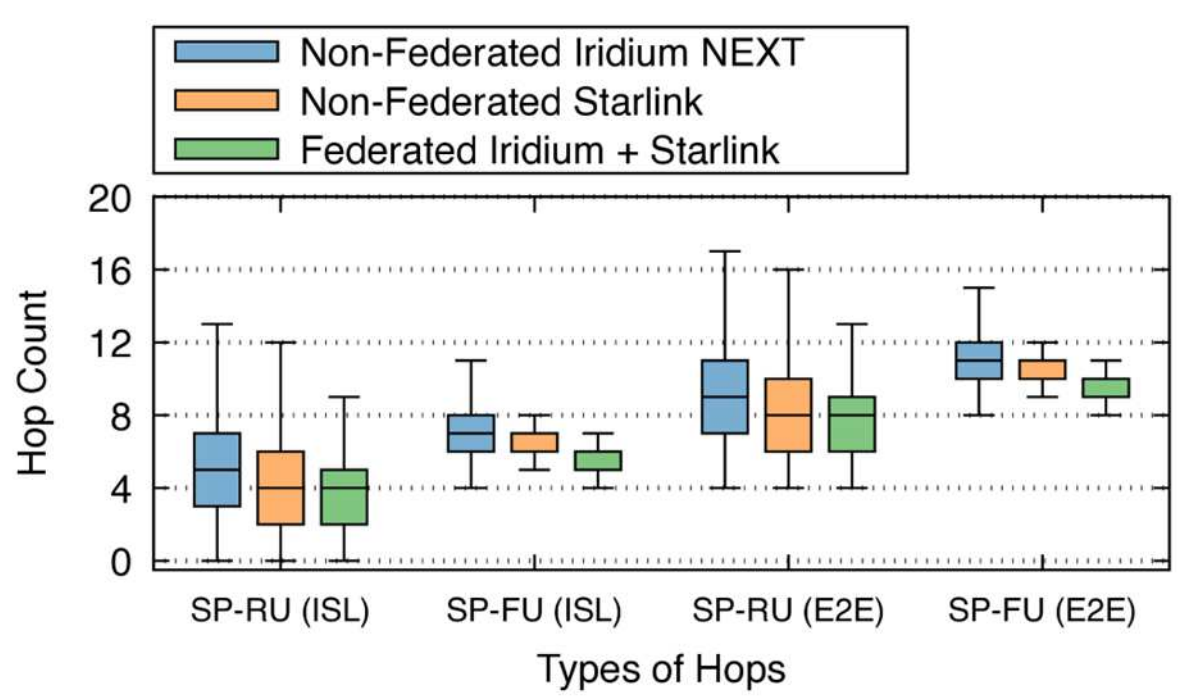

**FIGURE 9.** **Average hop counts for all architectures.**

Figure 10 presents the call block percentages for all three architectures. A *call block* is defined as the absence of a feasible inter-satellite routing path between the entry and exit satellites at the time of call initiation. A request drop occurs when the ISL graph is disconnected. The non-federated Iridium Next architecture experienced a significant number of call blocks (14.2% under SP-RU and 12.2% under SP-FU), primarily due to the absence of viable ISL paths. The non-federated Starlink architecture showed improved connectivity with a call block of less than 5%. Most notably, the federated architecture eliminates call blocking entirely under both SP-RU and SP-FU. This results in a 100% increase in connectivity compared with non-federated architectures (i.e., no call blocking in the federated case).

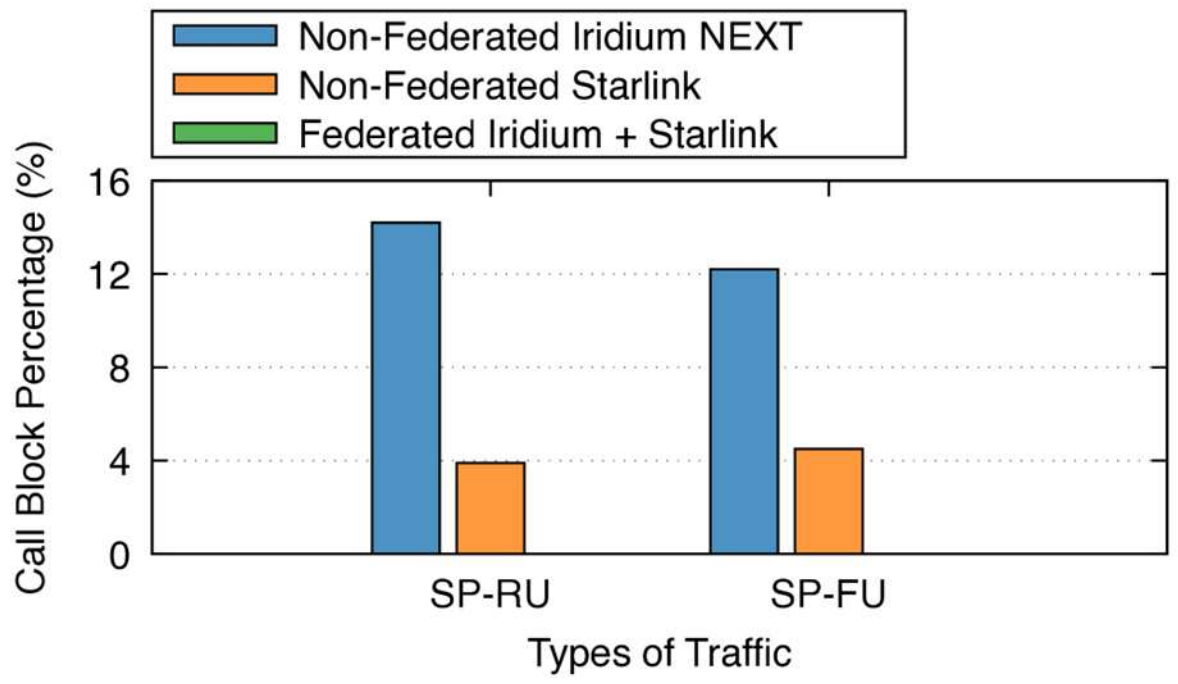

**FIGURE 10.** **Percentage of call block for all architectures.**

Figure 11 shows the Cumulative Distribution Function (CDF) of the E2E delays for SP-RU. The federated architecture demonstrated a clear leftward shift in distribution, indicating a reduction in the overall delay. For SP-RU, the maximum observed delay decreases from 231.24 ms (non-federated Iridium Next architecture) to 122.22 ms (federated Iridium-Starlink architecture), representing a nearly 47% reduction in worst-case latency.

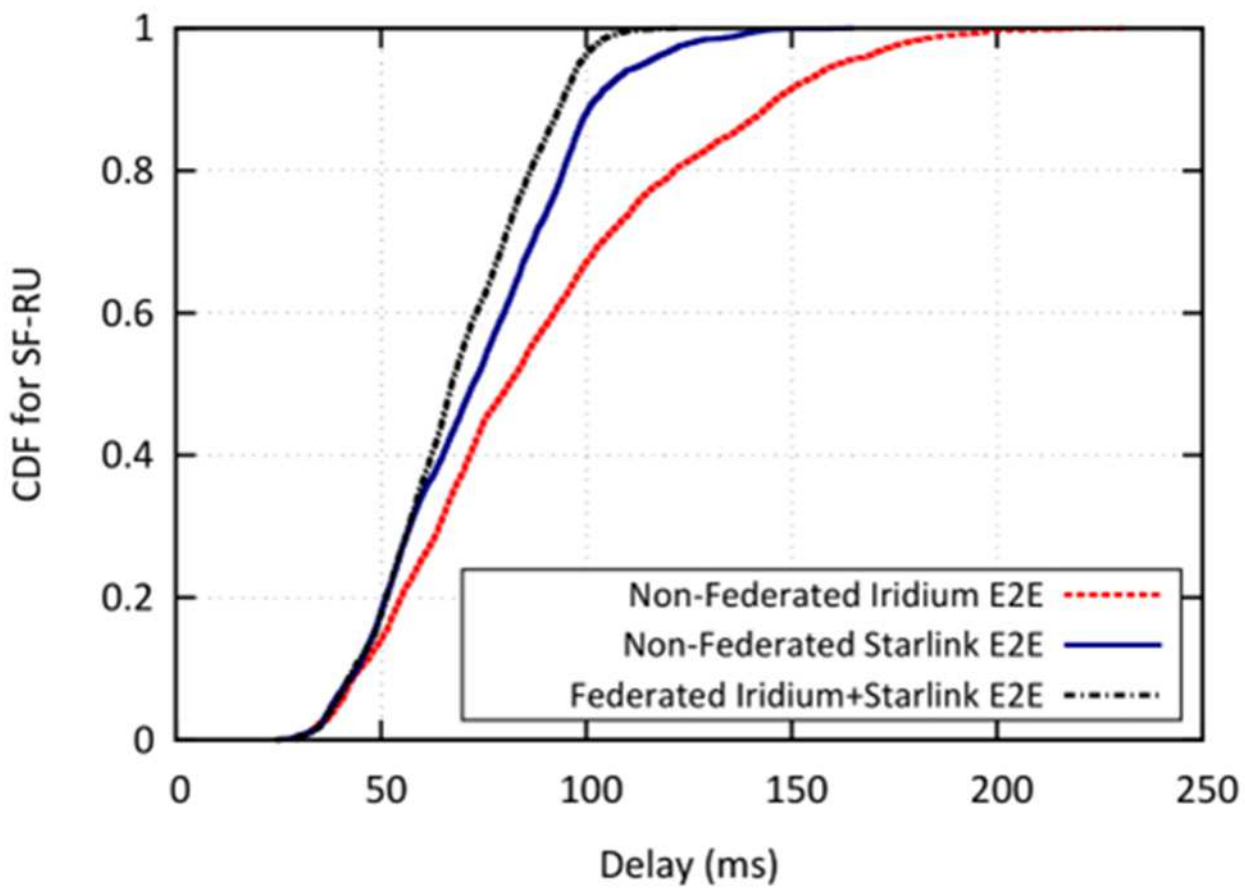

**FIGURE 11.** **SP-RU CDF for all architectures.**

Under SP-FU in Figure 12, the delay spread narrows considerably, reflecting improved stability. The federation not only reduces the mean delay but also suppresses extreme delay events, which are critical for real-time and mission-critical services.

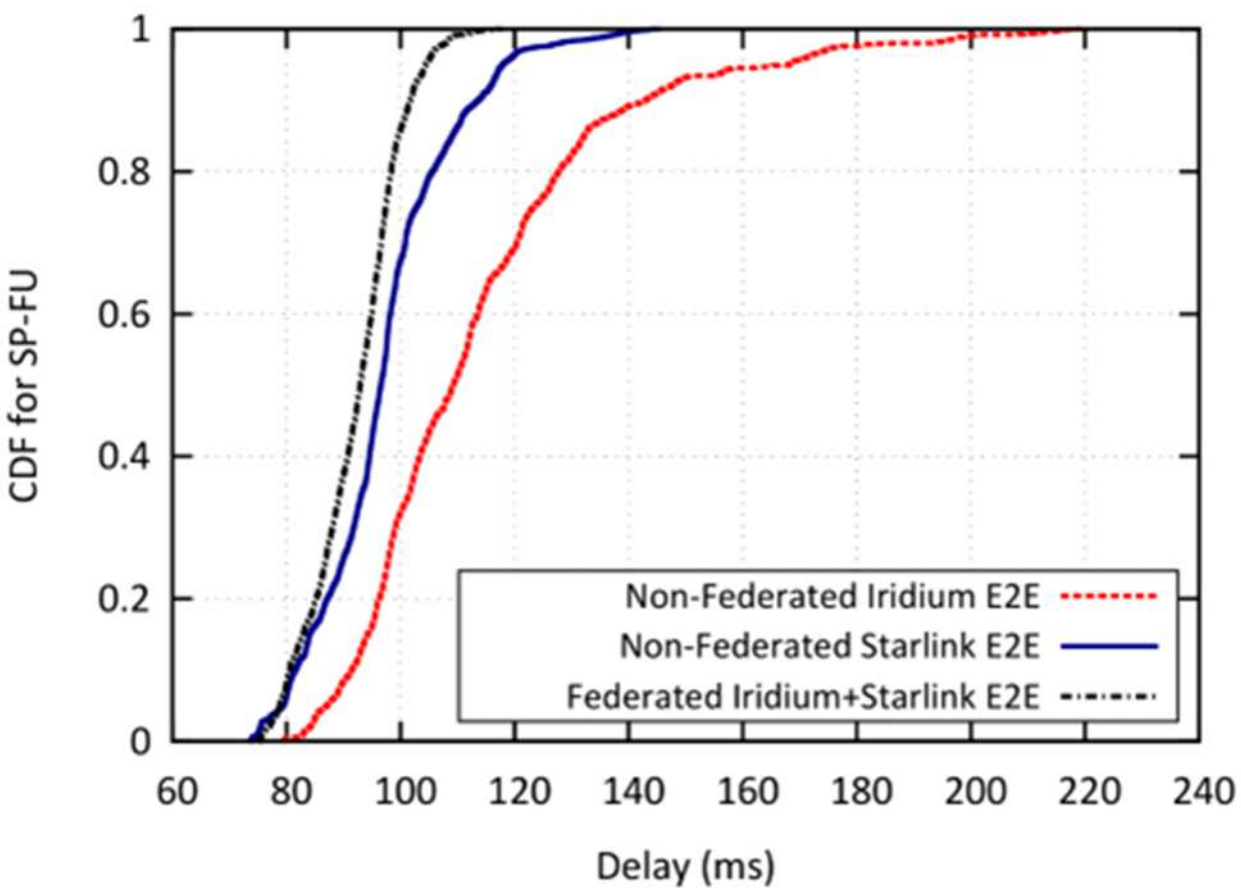

**FIGURE 12.** **SP-FU CDF for all architectures.**

Figure 13 shows the Coefficient of Variation (CV) of E2E delays, which quantifies latency stability. Under SP-RU, the non-federated Iridium Next architecture exhibits the highest variability (CV = 0.43), followed by the non-federated Starlink architecture (0.32) and the federated Iridium-Starlink architecture (0.27). Federation reduces the variability by approximately 37% compared with the non-federated Iridium Next architecture. However, Figure 13 shows that SP-FU produced significantly lower variability across all the architectures. For instance, the non-federated Iridium Next architecture's CV decreases from 0.43 to 0.21, and the federated Iridium-Starlink architecture achieves the lowest variability at 0.08. This indicates that although worst-case routing increases the average delay, it yields more deterministic behavior due to the constrained routing topology and reduced path randomness.

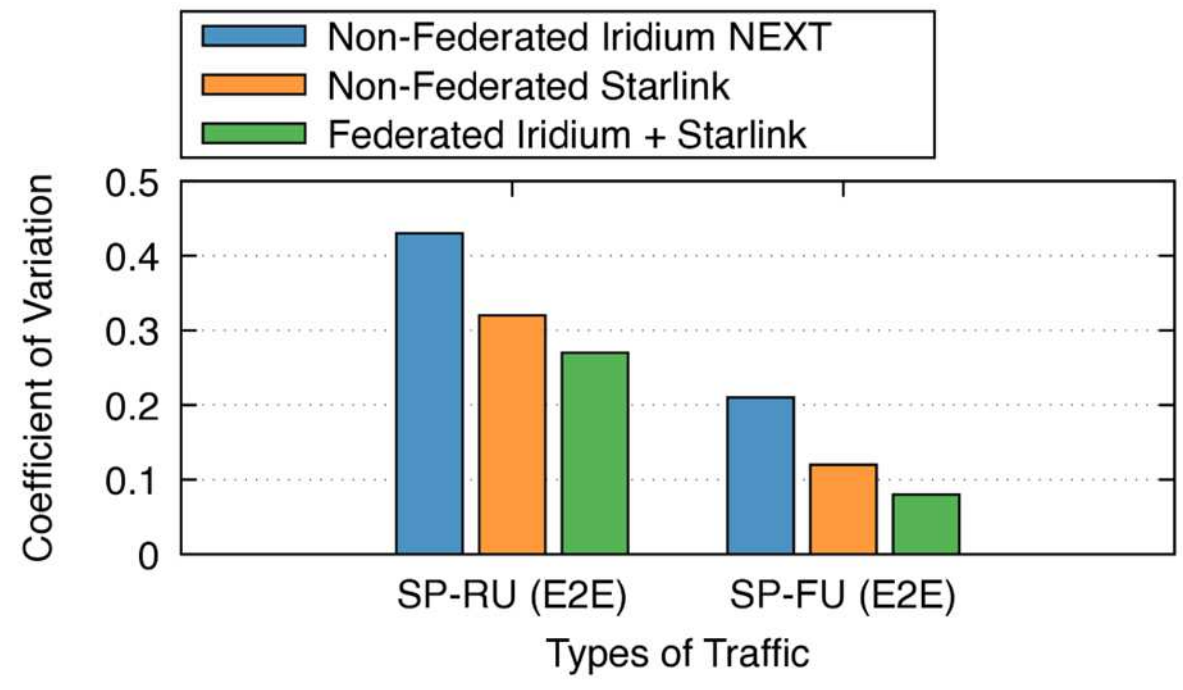

**FIGURE 13.** **SP-RU CDF for all architectures.**

### *D. OPEN RESEARCH PROBLEM FOR FEDERATION IN NTN SYSTEMS*

1. **Interface design:** The federation interface (Nsnf) should be standardized to enable interoperability between federated operators. It may expose its functionalities through a standardized Application Programming Interface (API), such as a RESTful API, together with the underlying communication protocols required to establish, manage, and terminate federation procedures between two or more operators. Capabilities can be:

   a. Continuous KPI/resource monitoring for the federated resources.
   b. Advertise and discover shareable resources (link capacity, ISLs, onboard computational resources, and gateways).
   c. Resource request and negotiation based on QoS requirements (e.g., latency, bandwidth, service duration, and computational capacity).
   d. Resource reservation, Virtual Network Function (VNF) initialization, routing changes, or slice configuration.
   e. Extend, modify, and tear down the federation connection
   f. Security.

2. **Network function integration:** Integration of the Synthesized Network Function (SNF) service chain with other Network Functions (NFs) in the network core, such as the Session Management Function (SMF) and UPF for packet routing and forwarding.
3. **Inter-operator federation:** If the satellite operator differs from that of a neighboring satellite operator, roaming/handover protocols, authentication/trust frameworks, and Service Level Agreement (SLA) coordination must be in place.
4. **Interoperability:** It can be achieved through unified control-plane APIs, common network function abstractions, and standardized service exposure across domains.

## VI. NEW ROUTING SCHEMES FOR NTN

Given the complexities of satellite constellation networks, including dynamically changing topologies, multi-orbit heterogeneity, and stringent QoS requirements, several novel routing approaches have been investigated. The following research directions are particularly noteworthy.

- **Broadband Support:** High-throughput applications (e.g., video streaming [34]) require congestion-aware routing. Combined with low-latency requirements for QoS demands of control services
- **Terrestrial Integration:** Seamless handovers for end users via cross-domain (TN&NTN) routing with native NTN integration are essential for uninterrupted service delivery.
- **Resilience:** The support of (post-quantum) secure protocols and resilience against Denial-of-Service (DoS) attacks, signal jamming, and node or link failures is critical. Path diversity and backup alternatives ensure robust routing.
- **AI Integration:** AI enhancements for real-time traffic prediction, anomaly detection, and adaptive path distributions are considered.

The dynamic topology of LEO constellations, driven by satellite motion, results in predictable yet frequent handover events. Consequently, routing protocols must dynamically adjust paths without incurring excessive signaling overheads. Novel solutions often rely on flexible and topology-aware approaches to locally resolve handovers and failures. Routing convergence is a key metric in this context, because centralized control can be affected by long propagation delays owing to the physical size of the ISL network.

Various strategies have been developed to address these challenges. To enable precise traffic engineering and flexibility, many proposed schemes rely on Segment Routing and/or Software-Defined Networking (SDN). Backup paths are considered to ensure resilience to failures and outages.

In the context of Internet Engineering Task Force (IETF) activities, an architecture is proposed to enable scalability by introducing network hierarchy [35]. This approach is based on an Intermediate System-to-Intermediate System (IS-IS) design, utilizing slices of the constellation that consist of neighboring orbital planes. Similar concepts have been explored using distributed SDN, in which scalability is achieved by subdividing the network into smaller domains with varying levels of autonomy [36]. Given the programmability and softwarization of the approach, a centralized master controller at a network control center can provide adapted routing logic [37].

Nevertheless, because the onboard satellite processing capabilities are limited, ground-based network management remains an essential component of most state-of-the-art designs. The placement of routing logic and control functions has also been discussed in 3GPP's proposed functional splits for 6G [38],[39]. A diverse set of architectures featuring various placements for core functionalities and distributed units is discussed. Complementing these developments is a trend toward AI-native networks. For routing, ML models, including Long Short-Term Memory (LSTM) networks and transformers, have been proposed to forecast traffic patterns and congestion hotspots, enabling pre-emptive re-routing. Graph Neural Networks (GNNs) are used to model constellation topologies and extract relevant features. Moreover, Deep Reinforcement Learning (DRL) agents are trained in simulated environments to develop adaptive policies for load balancing and fault recovery, thereby outperforming rule-based algorithms. Different architectures have been investigated, including hop- and path-based solutions [40].

Finally, designs for multi-layer and multi-orbit constellations (hybrid LEO-MEO-GEO and federated systems) demand further innovation [41]. It is crucial to find suitable paths across orbital layers based on the latency, reliability, and resource availability. Hierarchical control schemes tailored to a specific topology and coordinated by ground stations are particularly interesting. Coupled with AI-driven predictive analytics [42], these approaches will enable NTNs to evolve toward increasingly heterogeneous architectures.

### A. INSIGHTS FROM SYSTEM-LEVEL SIMULATIONS

In this section, a system-level simulator is adopted to validate key assumptions. The reference scenario shown in Figure 14 was based on previous investigations in this context [36]. The performance of the QoS-based routing scheme was evaluated in terms of QoS compliance, considering end-to-end latency and packet-dropping rate as the primary performance metrics. In Figure 15, the average end-to-end latency of sessions is compared across a shortest-path-first routing scheme, an ideal "God's Eye View" routing scheme, and a distributed, load-balancing routing scheme. The approaches and simulator environment were discussed in more detail in [36].

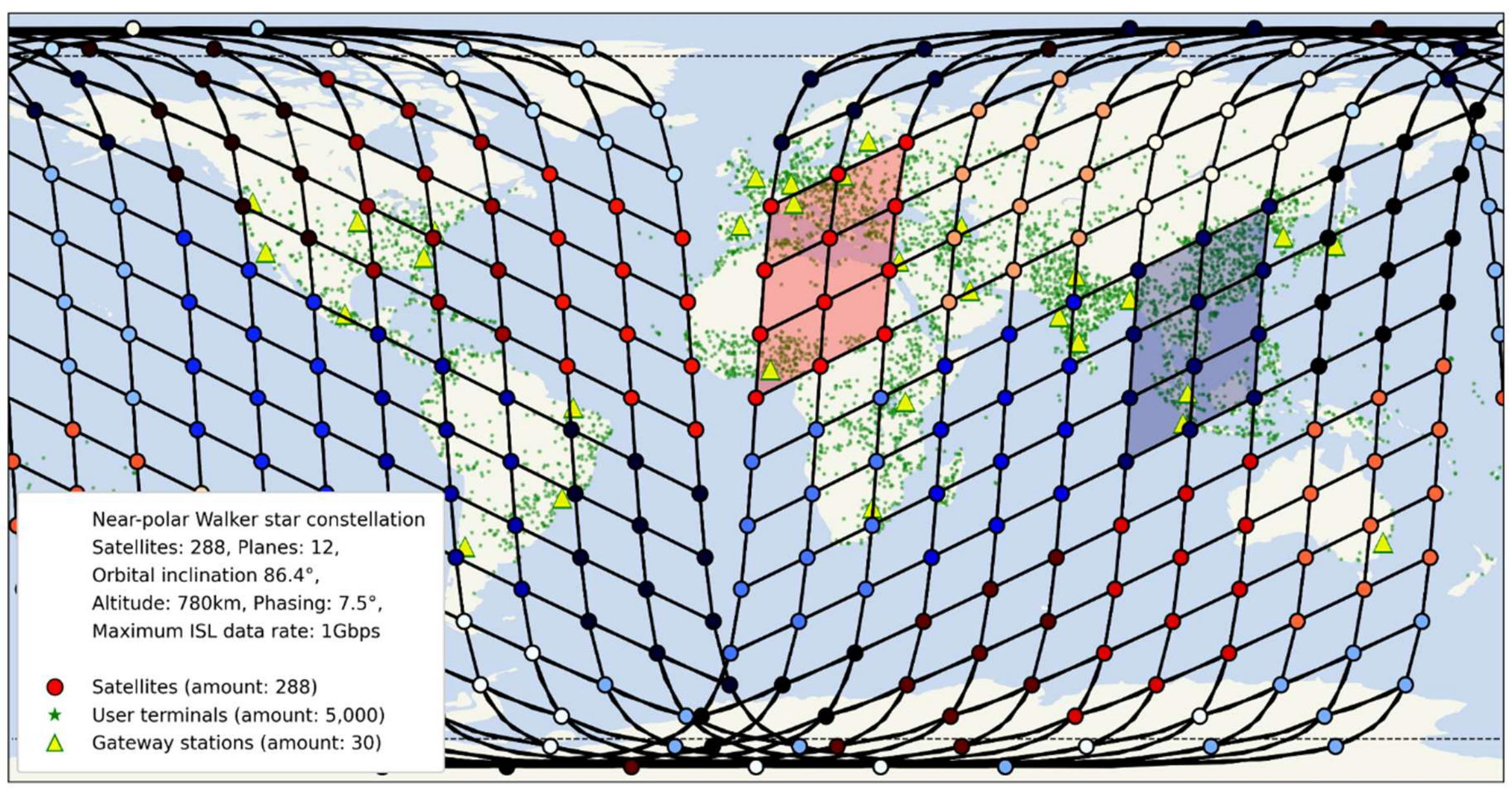

**FIGURE 14.** Reference scenario for routing analyses. Colored sets of satellites indicate exemplary clusters for the distributed routing scheme.

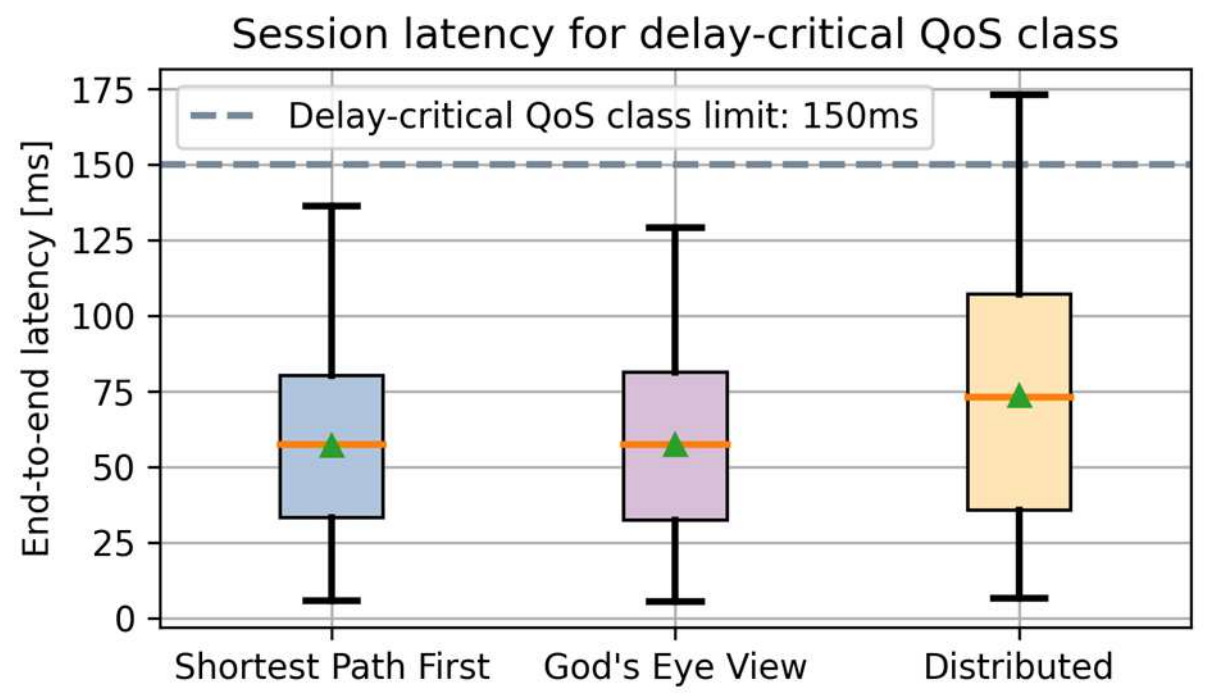

**FIGURE 15.** Comparison of the resulting average latency per session for different routing approaches. A delay-critical QoS class is assumed.

The comparison highlights that for delay-critical QoS classes, the end-to-end path centralized formulations (like for Shortest Path First and God's Eye View) achieve lower latency than distributed routing, whose local routing decisions may result in longer, zigzag paths that occasionally exceed the 150 ms latency constraint. Nevertheless, distributed routing remains attractive because it provides scalable and localized decision-making while maintaining acceptable latency performance. For QoS profiles with more relaxed latency requirements, where throughput becomes the primary performance objective, the load-balancing capability of the distributed routing scheme allows higher throughput than the shortest-path-first approach. The corresponding results are shown in Figure 16. Here, the gap between the network load and throughput of the Shortest Path First scheme indicates significant packet drops due to network congestion. A direct comparison of the packet drop rates is shown in Figure 17. Here, a different QoS class is considered, namely a best-effort service characterized by relaxed latency requirements but stringent packet dropping rate constraints. While the Shortest Path First approach is prone to congestion, leading to high packet drop rates, the load-balancing Distributed scheme can meet the permitted rates (drop rates below $10^{-6}$). Even with this high traffic volume, the theoretical God's-Eye-View routing benchmark can completely avoid packet drops due to its immediate, complete knowledge of the network. These results highlight the advantages of informed and rapid decision-making for routing in NTNs.

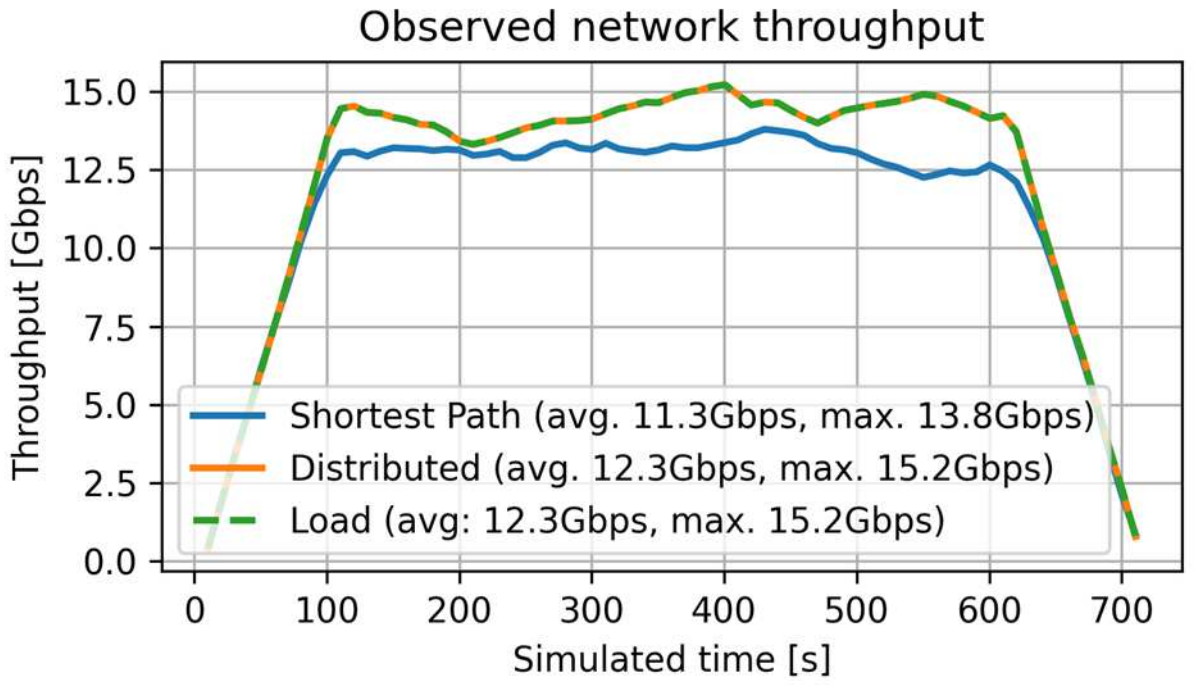

**FIGURE 16. Observed throughput for two routing approaches (i.e., Shortest Path First and Distributed schemes) together with the offered network load.**

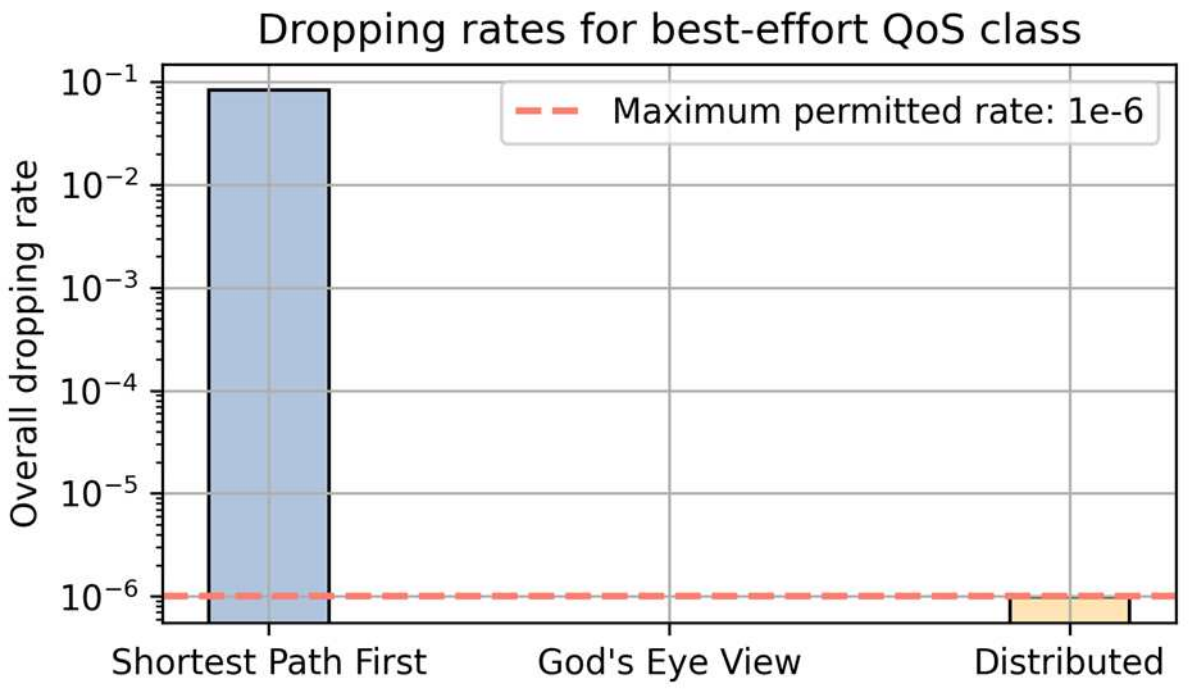

**FIGURE 17. Comparison of the resulting packet drop rates for different routing approaches. A best-effort QoS class is assumed.**

## VII. EDGE INTELLIGENCE

Recent advances in Orbital Data Centers (ODCs) have moved from concept to early implementation, with major industry players such as NVIDIA and AWS exploring space-based infrastructures to support AI workloads directly in orbit. Notably, this trend is supported not only by datacenter operators/providers but also by space constellation operators such as Starlink, which has recently provided more details about its plan to deploy very large constellations serving as ODCs in space. These initiatives aim to enable in-orbit data processing, storage, and analytics, reducing dependence on terrestrial facilities while exploiting advantages such as abundant solar energy and passive cooling, as highlighted in recent studies and pilot concepts. As a result, ODCs are emerging as key enablers of future space-edge ecosystems, tightly integrated with satellite networks to support low-latency, data-intensive applications and advanced AI services [43],[44]. Nevertheless, the potential of ODC in space presents important challenges for *heat dissipation* and *downlink data rate*. Regarding the former, heat dissipation in space can be complicated because the vacuum nature of space implies that dissipation must occur via infrared electromagnetic waves, that is, thermal radiation. Although radiated power scales with the fourth power of the radiator temperature under the Stefan–Boltzmann law, the allowable operating temperature of onboard electronics limits the maximum radiator temperature. Consequently, the large thermal loads generated by ODCs require extensive radiator surfaces, posing significant Size, Weight, and Power (SWaP) challenges for satellite platforms [45]. As for the second challenge, lessons learned from terrestrial data centers indicate that high data rates must be provisioned to make large volumes of data available for processing and to deliver the final results to end users. In this sense, the available data rate remains limited compared to terrestrial counterparts, unless free-space optics is used in very specialized satellite infrastructures capable of coping with FSO signal blockage from severe weather conditions (e.g., clouds). Therefore, ODC, although a very attractive and intriguing area of research, requires further investigation.

Edge Intelligence (EI) represents the convergence of Edge Computing (EC) and AI, establishing a paradigm in which intelligent data processing and decision-making occur near the data source [46]. In the context of 6G and integrated Terrestrial and Non-Terrestrial Networks (T-NTNs), EI is a foundational technology that equips edge nodes, such as satellites, HAPs, and Low-Altitude Platforms (LAPs), with the dual capability to perform AI tasks locally and serve as distributed platforms for offloading computation when needed. This requires a framework in which processing tasks are intelligently distributed across the layers of the T-NTN architecture to satisfy diverse service requirements [47],[48]. The unique characteristics of NTNs, such as long propagation delays, constrained bandwidth, and the need for autonomous operation, render EI beneficial and essential. By processing data closer to their source on a satellite or aerial platform, EI significantly reduces the latency inherent in round-trip proèagation with ground stations, which is a critical factor for real-time applications [47]. This localized processing enhances bandwidth efficiency, as only relevant information or model updates are transmitted rather than vast amounts of raw data. This approach inherently improves data privacy and security by maintaining sensitive information on-orbit, and contributes to the scalability and reliability of the network, particularly when ground connectivity is intermittent [47],[49].

The realization of EI in T-NTNs is supported by flexible, virtualized, and intelligent network architectures. The key among these enablers is the O-RAN, which provides a standardized framework for deploying virtualized and intelligent RAN functions across a distributed T-NTN infrastructure [50]. The O-RAN's distributed control, facilitated by Non-Real-Time RIC and Near-Real-Time RIC, enables intelligent optimization across different time scales, creating a structured environment for managing ML models [50]. Network Slicing complements this scenario by enabling the logical partitioning of physical network resources into multiple virtual slices, allowing diverse EI services to be delivered simultaneously with customized performance guarantees [47].

Distributed Learning (DL) forms the core of this intelligence, allowing multiple nodes to collaboratively train models without centralizing raw data. This paradigm includes several powerful techniques that are suited to the NTN environment. In particular, Federated Learning (FL) enables collaborative training while preserving data privacy by exchanging model updates instead of raw data [51]. For resource-constrained nodes, such as UAVs, Split Learning (SL) partitions a model between the device and server, whereas Transfer Learning (TL) accelerates training by leveraging pre-trained models [49]. To address the significant data heterogeneity across NTN nodes, Personalized FL (PFL) tailors models to local data, and Decentralized FL (DFL) facilitates peer-to-peer model sharing via ISLs, removing the need for a central server [52].

The synergistic combination of these concepts has led to the proposal of sophisticated EI frameworks tailored for specific NTN applications. For instance, the Distributed Learning-as-a-Service (DLaaS) framework uses network slicing to deploy chains of virtualized learning functions across T-NTN layers, serving diverse Internet of Vehicles (IoV) applications [47]. For Earth Observation (EO) missions, the Generalized Federated Split Transfer Learning (GFSTL) framework proposes a hierarchical architecture in which UAVs capture high-resolution imagery, HAPs act as intermediate FSL servers, and LEO satellites serve as main FL servers, combining the strengths of FL, SL, and TL for enhanced accuracy and efficiency [49]. The same approach has also been extended to Intelligent Transportation Systems, where GFSTL not only achieves faster convergence and higher detection accuracy but also substantially reduces communication overhead relative to baseline FL, while maintaining both detection accuracy and end-to-end latency invariant as the number of participating users increases [53]. To address data heterogeneity in LEO constellations, the ALANINE framework first employs DFL to collaboratively train an onboard image Super Resolution (SR) model that enhances the quality of Earth observation images [52]. It then applies PFL with model pruning to generate lightweight, satellite-specific models adapted to local data characteristics.

Despite these advantages, significant challenges remain. The dynamic and resource-constrained nature of NTN nodes makes the optimal placement of learning functions and resource allocation highly complex [47],[54]. Furthermore, data heterogeneity and model drift can compromise the performance of distributed models [49],[52]. Managing EI in high-speed mobile environments while mitigating the new security vulnerabilities introduced by the distributed architecture is a crucial area for ongoing research to unlock the full potential of EI for autonomous, efficient, and green NTNs.

## VIII. GREEN NETWORKING CONCEPTS FOR NTN

Energy awareness and, more generally, sustainability have become major concerns across all areas of computing and networking. This trend is even more evident in the evolution toward 6G, where energy efficiency is recognized as key design objective rather than a secondary optimization criterion (see, e.g., Section 2.5 in [55], which states that "…the selection of suitable mechanisms ought to include an *energy consumption* KPI at the same level of today's focus on performance KPIs such as *throughput or delay*."). In fact, these two KPI categories may be in conflict; therefore, it is often necessary to consider the trade-offs between them, particularly when defining optimization objectives.

Three phases can be identified in management and control mechanisms aimed at achieving a trade-off between network performance and energy consumption, in either TN or NTN.

1. Measurement and data collection
2. Synthesis of a control/management strategy
3. Actuation of the ensuing decisions.

Regarding points 1 and 3 above, it is notable that although many opportunities exist at various operational levels and time scales to monitor and influence performance-related KPIs, the same cannot be said for energy-consumption-related KPIs. In this respect, the definition of APIs capable of conveying information on energy consumption and enabling its dynamic management (possibly over different time scales) has a long history in computing systems. A notable example is the Advanced Configuration and Power Interface (ACPI) [56], which provides a standardized interface between the hardware and software layers to support energy-aware operating states. However, comparable functionality was not introduced into the networking domain until 2013.

ACPI introduced the concepts of Power States (C-states, $C_x$, $x = 0, 1, \ldots, X$) and Performance States (P-states, $P_y$, $y = 0, \ldots, Y$), which can be individually configured and tuned for each processor core. $C_0$ is the active power state; the states from $C_1$ onward are processor sleeping or idle states (in which the processor consumes less power and dissipates less heat). In the $C_0$ state, the ACPI allows the core to adapt its processing rate (Adaptive Rate, AR) by selecting different P-states that either alter the operating frequency and/or voltage or throttle the clock, thereby imposing a trade-off between performance and energy consumption. On the other hand, Low Power Idle (LPI) mechanisms enabled by C-states also entail a similar trade-off owing to the performance degradation introduced by longer wake-up times associated with "deeper" sleep. The combined effect of AR and LPI on energy consumption and performance requires careful tuning [57].

With regard to the extension of these concepts to telecommunication systems and networks, following inputs from the ECONET project (see [58]), the European

Telecommunications Standards Institute (ETSI) defined the Green Abstraction Layer (GAL) standard [59], which introduces Energy Aware States (EASs) with characteristics resembling those of the ACPI, along with a multi-layered abstraction interface for the hardware and physical resources and their Power Management Primitives (PMPs). However, it is worth noting that the further extension of these GAL concepts to the current softwarized, virtualized, and integrated networking environments is not immediate, as in legacy networking devices, access to hardware (which is the ultimate source of energy consumption) was directly managed by a single functionality, rather than occurring in a multi-tenant context of Network Functions Virtualization (NFV), where hardware resources are shared and accessed through the mediation of various software layers. In this framework, there are three main causes of the separation between the Virtual Network Functions (VNFs) and the hardware executing them, which can be identified as follows:

- The execution of multiple concurrent VNFs is mediated either by a hypervisor and its scheduling policies in the case of Virtual Machines (VMs), or by an orchestration and management platform, such as Kubernetes, in the case of containers.
- Multiple tenants sharing the same hardware may be present;
- It is difficult to establish a direct correspondence between the use of virtual resources (e.g., vCPUs and virtual Central Processing Units) and the EASs of the hardware hosting them.

Despite these challenges, which require further investigation, both ETSI and ITU have considered these issues by defining EASs specific to VNFs and Network Services, and by providing the specification of the interfaces for all reference points of the NFV ETSI MANO architectural framework [60], along with the operations of provisioning, release, and monitoring, in the new GALv2 standard [61].

When designing and implementing control and management strategies, the trade-off between performance and energy-efficient KPIs should always be considered, even when dealing with VNFs. Indeed, although virtualization can greatly increase flexibility in resource allocation and foster energy saving by allowing the dynamic consolidation and migration of functionalities, the use of general-purpose servers may entail greater energy consumption with respect to that of customized hardware (see, e.g., [62]), which may be further exacerbated by the acceleration hardware adopted to carry out computational tasks connected with the execution of Artificial Intelligence/Machine Learning (AI/ML) algorithms. We believe that Dynamic Adaptation strategies, such as LPI and AR, can still play a role in this context by leveraging GALv2's capabilities to improve energy-efficiency KPIs. In this context, the ongoing Horizon Europe 6Green project [63] adopted three lines of action:

- **Edge Agility** provides smart, fast, and automated horizontal scalability to vertical applications and related slices across the 5/6G edge-cloud continuum.
- **Green Elasticity** dynamically and adaptively provides energy-aware hardware-assisted acceleration for network functions and vertical applications, thereby enabling smart vertical scalability.
- **Energy-Aware Backpressure** introduces a set of cross-domain observability mechanisms and analytics to evaluate the energy and carbon footprint of a vertical application, slice, or overall 5/6G network imposed on the edge cloud infrastructure.

In O-RAN for NTN, the allocation of DUs, CUs, and RICs, whether in the sky or on the ground, can significantly affect energy consumption for computation and/or communication. The allocation of rApps (for Non-Real-Time RICs), xApps (for Near-Real-Time RICs), and dApps (for Real-Time sub-millisecond functionalities) can be dynamically optimized between TNs and NTNs.

In cloud-native and NFV settings within integrated terrestrial and wireless networks (including NTNs), orchestration is necessary to address the challenges of decentralization. In the application domain, services are designed and implemented as chains of microservices with specific communication requirements; through intent-based networking, these requirements can be conveyed to the networking domain to create Network Services composed of VNFs. In this framework, the concept of *separation of concerns* between the cloud-native domain of vertical applications and the telco domain of network functions and services has been recognized with the creation of separate orchestrators: a Network Applications Orchestrator (NAO) interacting with the MANO orchestrator through the mediation of an Operations Support System (OSS) [64]. Given the increasing number of satellites deployed in LEO constellations, and the hierarchical setting of NTNs, where GEO, MEO, and LEO satellites may interact among them, with HAPs and UAVs, as well as with Mobile Edge Computing (MEC) devices in the network edge, a similar separation of concerns has been suggested to deal with the increased complexity of the networking domain. In this case, an NTN Orchestrator might be worth introducing to handle both performance and energy efficiency KPIs to alleviate the task of the Network Functions Virtual Orchestrator (NFVO) dealing with TN functionalities (possibly through the mediation of the NTN Network Control Center(s)/Satellite Gateways; see Figure 8 in [65] and Figures 15 and 16 in [66]).

Within the framework of NTNs, the full on-board Base Station (BS) functionality and functional-split option for regenerative satellite payloads, as considered by 3GPP [67], are of particular interest for investigating their energy-efficiency implications. The SMARTEN6G project [68], funded by Xjenza Malta, investigated some of these aspects. By exploiting a Digital Twin implementation and, where possible, in-field measurements, the evaluation of energy consumption should focus on both computational and communication efforts. In particular, the presence of full BS functionality is likely to entail greater computational effort than hosting only DU functionality; however, the communication load should also be considered. In the full BS case, the Satellite Radio Interface (SRI) carries the NG interface between the onboard BS and the 5G Core, whereas in the DU/CU split case, SRI carries the F1 interface between the DU and the CU (see Section IV). Another point of interest in this context is the placement of Near-Real-Time RIC onboard or in the cloud. The convenience of placement may depend on the application's specific functionality. As noted above, a highly relevant research problem concerns the dynamic placement of applications (xApps and dApps) in the tradeoff between responsiveness and energy consumption. Another relevant aspect of the functionality of Near-Real-Time RIC is its potential to dynamically determine "the optimal functional split based on the collected network status data and redeploying the network functions in the CU and DU according to it," as noted in [69]. When deploying AI/ML algorithms in Near-Real-Time RICs, their energy requirements should be carefully evaluated, particularly for components operating onboard the NTN segment. In general, onboard AI functions should preferably be limited to lightweight inference and control tasks based on parameterized models (e.g., neural networks), whose parameters have been obtained through training over longer timescales. Such computationally intensive training processes are more naturally executed by cloud-based Non-Real-Time RICs. Nevertheless, NTN infrastructures may also be exploited for computation offloading beyond traditional application-level services. For example, in federated learning, parts of the training process associated with control algorithms could also be distributed to NTN nodes, allowing satellites to actively participate in collaborative model training. A notable recent proposal in [70] presents an AI-RAN orchestrator that fosters the convergence of O-RAN and AI-RAN architectures.

We present preliminary results from the SMARTEN6G project on energy-efficiency measurements. SMARTEN6G considers two use cases: Vehicle-to-Everything (V2X) and Augmented Reality/Mixed Reality (AR/MR). Figure 18 [71] highlights some details of the SMARTEN6G Digital Twin testbed for measuring the power and energy consumption of the UE and the regenerative satellite. The complete testbed, when used to study V2X and AR/MR applications, consists of a UE connected to the regenerative satellite edge-AI payload via the Digital Twin Channel Emulator.

The results reported here pertain to a simpler scenario used to evaluate the power consumption of an edge-AI application on a simulated regenerative LEO NTN satellite. The simulation platform is based on OpenAirInterface (OAI) and its RFSim channel simulator, which together emulate a regenerative monolithic gNB onboard the satellite [72],[73].

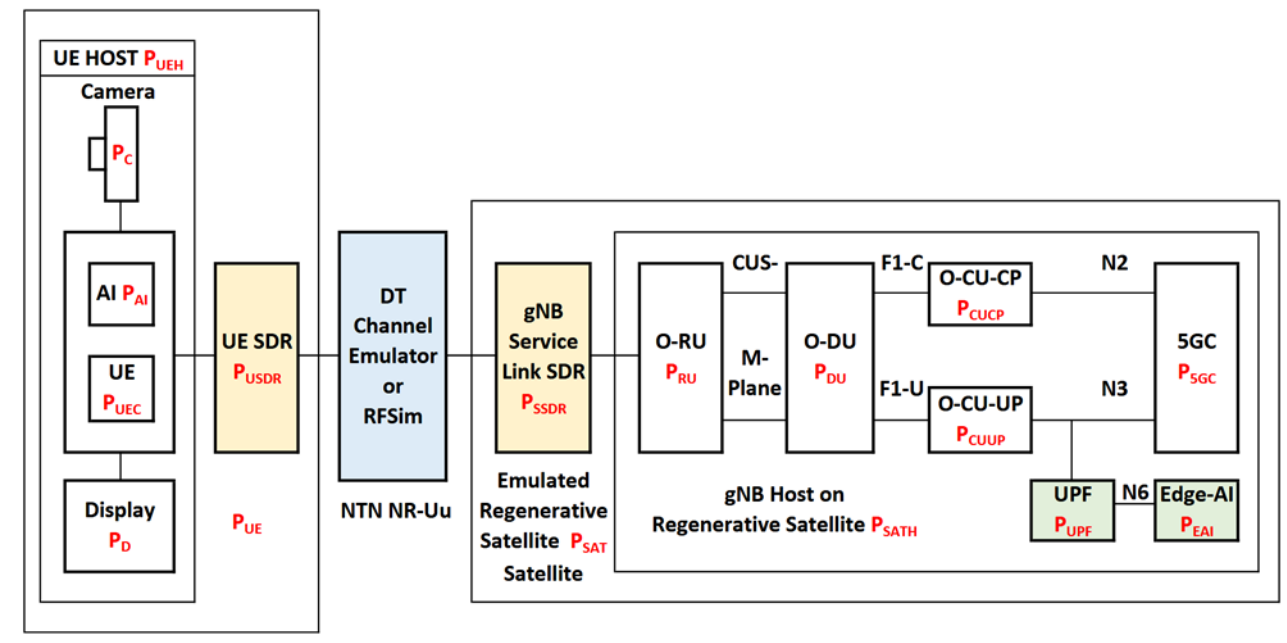

**FIGURE 18. SMARTEN6G block diagram for measuring the power and energy consumption in AI/ML task offloading to edge-AI flying component of NTN.**

The edge-AI application consists of a UE that offloads a video stream captured from a vehicle to a YOLOv8n-based car-detection program running on the simulated LEO NTN satellite. The video, captured from an onboard webcam, is transmitted at 1 Mbit/s to the satellite payload, which performs real-time vehicle detection. The satellite payload is emulated on an ASUS Revel Canyon NUC 14 Pro (Tall), a compact mini-PC powered by the Intel Core Ultra 7 155H (Meteor Lake) processor, whose CPU cores are hereafter referred to simply as the CPU. In particular, this processor integrates a heterogeneous CPU comprising 16 cores (6 Performance, 8 Efficient, and 2 Low-Power Efficient cores), an Intel Arc integrated GPU (8 Xe cores @ ≈2.25 GHz), and an Intel AI Boost Neural Processing Unit (NPU), making it an energy-efficient platform for edge-AI and multimedia tasks. The YOLOv8n vehicle-detection script, written in Python, exploits the Core Ultra 7 CPU, Intel Arc GPU, and NPU via Intel's OpenVINO (Open Visual Inference and Neural Network Optimization) Software Development Kit (SDK) [76].

The NUC mini-PC runs the OAI UE, gNB, and 5G core network elements [such as Access and Mobility Management Function (AMF), SMF, UPF, external data networks, and MySQL database] in Docker containers [72]. The RFSim module was executed on both the UE and gNB sides to simulate the LEO channel [73]. The uploaded video was routed through the UPF to the Python application, which was executed sequentially on the CPU, the GPU, and the NPU. Intel Running Average Power Limit (RAPL) counters (core, uncore, package-0, and psys) were used to derive the component-level power consumption of the CPU, GPU, NPU, and other subsystems [77]. Docker stats provide per-container CPU and memory utilization, which can be correlated with RAPL-derived power measurements to estimate the container-level energy consumption. Alternatively, a dedicated power logger can collect both the RAPL counters and external physical power measurements. An external Raritan PX3-5440V-M5K1 Power Distribution Unit (PDU) measures the physical power drawn by the NUC and serves as a ground-truth reference for validating the internal power estimates. A hash script automates the experiment: starting from a powered-down state, it sequentially launches the 5G core, gNB, and UE, streams video from the UE to the YOLOv8n application, and executes workloads on the CPU, GPU, and NPU. When the cycle is completed, the streaming and AI workloads, followed by the UE, gNB, and core, are shut down in the reverse order, returning the NUC to its baseline state. As the individual power contributions of the UE and 5G core are measured, they can be subtracted from the total NUC power to isolate the power consumption of the NTN 5G and edge-AI satellite payloads. Power consumption, as measured by the RAPL counters and attributed to Docker containers via Docker stats, is shown in Figure 19.

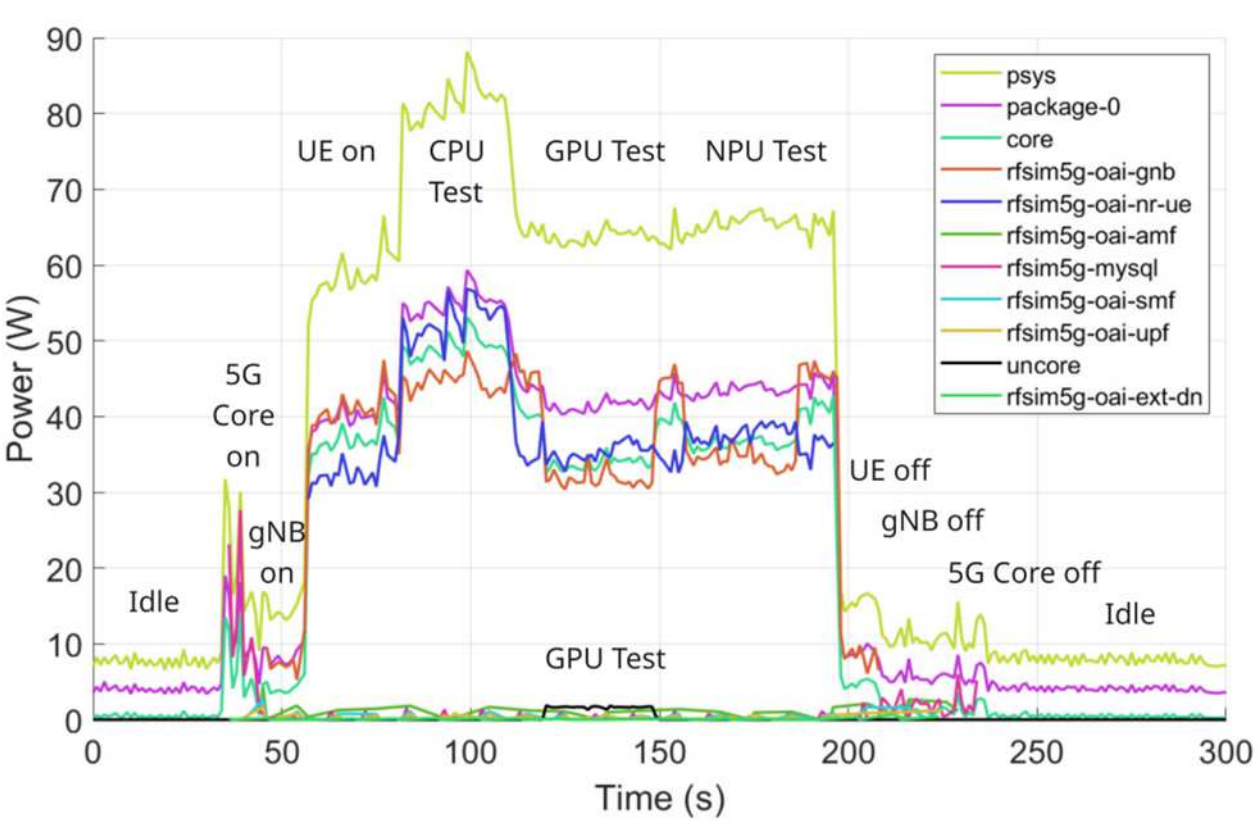

**FIGURE 19. Power consumption measured using RAPL and Docker stats attribution during the edge-AI offloading cycle.**

The power consumption measured by the power logger and the Raritan unit is shown in Figure 20. As shown in this figure, when the 5G core is powered on, an initial transient lasting approximately 10 s is observed, which is characterized by elevated power consumption. After this start-up phase, the power drawn by the 5G core containers stabilizes at a marginal level. When the gNB was subsequently activated, the total power increased by approximately 5 W, and then increased to approximately 60 W once the UE became active. This indicates that the power-intensive lower layer processing functions of the gNB are primarily engaged when the UE is connected.

During the video offloading and inference stage, when the YOLOv8n application was executed, the CPU power consumption increased by approximately 20 W. Offloading the same inference task to the GPU and NPU of the Intel Meteor Lake SoC reduced the AI processing power by approximately 15 W, demonstrating the benefits of hardware-accelerated energy-efficient computation for edge-AI workloads.

Figure 20 shows that the total power reported by the RAPL is consistently higher than that obtained from the Raritan PDU, confirming that the RAPL tends to overestimate the absolute system power compared to the calibrated external (ground-truth) meter.

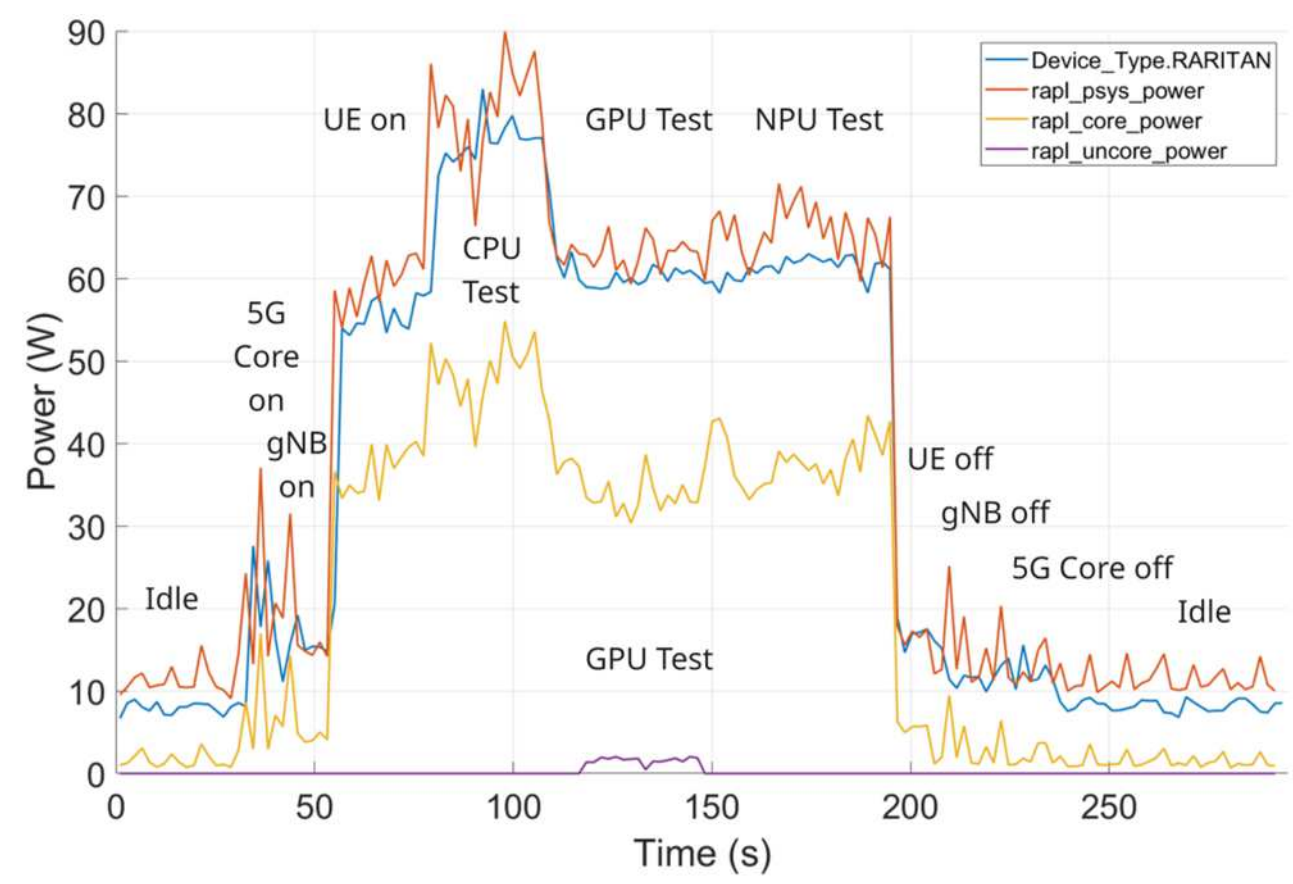

**FIGURE 20. Power consumption measured using a power logger and a Raritan PDU.**

## IX. SECURITY, PRIVACY, AND RESILIENCE IN FEDERATED T-NTN

Integrated TN-NTN 6G networks expand the attack surface across multiple domains, including the ground infrastructure, aerial platforms, and multi-orbit constellations. Adversaries may target satellite control links, gateway infrastructure, inter-satellite links, or edge computing nodes to degrade availability, compromise integrity, or exfiltrate data. Beyond malicious attacks, large-scale LEO constellations are also exposed to space-weather events, collision risks, and cascading failures, which can have a systemic impact on connectivity and safety-critical applications [78].

Hence, security, privacy, and resilience must be treated as critical design objectives for integrated T-NTNs [79],[80]. This includes secure-by-design architectures, formalized threat models, and continuous verification and monitoring, spanning the TN, NTN, and edge domains. Let us recall that the European DNA explicitly links resilience, preparedness, and security obligations across terrestrial and non-terrestrial infrastructure, further underlining the need for holistic security frameworks in future NTN deployments.

First, federation dramatically changes the trust model, where cross-operator interfaces, SLAs, and policy exchanges must be secured and monitored. A security architecture and trust model must be clearly defined for federated T-NTNs, outlining the trust relationships between operators, domains, and network functions. Specifically, a multi-operator federation typically implies that user traffic, control signaling, and management plane operations may transit through domains with non-uniform security. A zero-trust security model may be appropriate here, in which no domain, node, or link is implicitly trusted, and each access or transaction is continuously authenticated, authorized, and monitored. Key security architecture capabilities to consider include: (*i*) Strong, lifecycle-managed identities and credentials for satellites, gateways, RAN nodes, and edge computing entities; (*ii*) End-to-end confidentiality and integrity protection for user and control plane traffic with post-quantum-ready key exchange mechanisms for long-lived satellite assets; (*iii*) Cross-domain policy and telemetry exchange mechanisms allow operators to share security-relevant information, while preserving confidentiality and commercial constraints.

Second, the capabilities of the T-NTN system, such as federation and routing (as discussed in Sections V and VI), can be extended to incorporate security and resilience metrics as well as latency and capacity. In addition to propagation and transmission delays, link weights can encode risk scores, trust levels, or regulatory constraints (e.g., avoidance of specific jurisdictions for certain traffic classes) [81]. For example, the set of feasible paths in a federated network can be further constrained by operator-defined security policies such that only links and satellites with verified software baselines and acceptable risk profiles are used for critical services. Resilience can be enhanced by explicitly engineering the path diversity across operators, orbits, and technologies. Disjoint or partially disjoint paths can be pre-computed to protect against single-point failures, targeted jamming, or localized cyber incidents. In addition, routing controllers can leverage security telemetry (e.g., anomaly scores, detected intrusions, or jamming indicators) to dynamically steer traffic away from compromised assets, combining the "God's Eye View" global perspective with local, fast-reroute mechanisms for time-critical services.

Third, recognizing how incorporating edge intelligence introduces new risks motivates the use of models that support risk management and reduction. For example, federated and distributed learning can satisfy certain privacy requirements by exchanging only model updates and gradients. However, such models remain vulnerable to poisoning, inference attacks, and injection of backdoored models [82],[83]. Mitigation strategies include robust Byzantine-resilient aggregation of model updates, secure aggregation protocols that conceal individual contributions, and anomaly-detection techniques to identify malicious or anomalous updates. Remote attestation and integrity verification of EI components, combined with signed and versioned model artifacts, can help ensure that only vetted models are executed on satellites and edge nodes. Privacy can be further enhanced by leveraging differential privacy mechanisms and compression and sparsification techniques that limit the information content exposed during model exchanges [84],[85].

Moreover, security-energy efficiency tradeoffs have become nontrivial, as energy efficiency is a key objective in future NTN systems, particularly for space-borne and autonomous platforms. However, aggressive energy-saving strategies (e.g., deactivating monitoring functions, reducing cryptographic key sizes, or batching security logs) can unintentionally degrade the security posture or incident detectability. Conversely, strong cryptography, continuous telemetry collection, and frequent model updates incur nontrivial energy and bandwidth overheads. Therefore, orchestration frameworks for T-NTNs should jointly optimize the performance, energy, and security KPIs. For instance, orchestration logic can adapt the placement of security functions (e.g., intrusion detection, anomaly analytics, or model verification) between the TN and NTN segments based on current energy budgets, threat levels, and mission criticality. GALv2-like abstractions and NTN-specific orchestrators can expose security-relevant energy states and capabilities, enabling verticals to express intents that balance latency, reliability, energy, and security requirements in a unified manner.

However, several open issues remain in the realization of secure and resilient federated T-NTNs. These include the definition of interoperable, machine-readable security policies and SLAs across operators; scalable, privacy-preserving telemetry sharing for joint threat detection; and robust AI-enabled defense mechanisms capable of operating under intermittent connectivity and constrained resources. In addition, the long-life cycles of satellites raise questions about crypto-agility and long-term resilience to emerging threats, such as large-scale quantum computing. Future work should also consider formal methods and digital-twin-based validation of T-NTN security and resilience strategies, integrating realistic models of space-weather, orbital dynamics, and adversarial behavior. This is essential for providing quantitative assurance for safety and mission-critical services that rely on integrated terrestrial and non-terrestrial networks.

### A. EXPERIMENTAL RESULTS: CLOSED-LOOP AUTOMATION

As 6G networks become increasingly virtualized and distributed across multiple administrative and technological domains, the attack surface expands considerably, rendering them more vulnerable to performance overloads and DoS attacks. These issues are particularly important in mission-critical scenarios such as emergency response and public safety, where communication reliability is essential. In such contexts, disruptions in either the control plane or the user plane can propagate across federated domains, potentially affecting both TN and NTN. The following study addresses these challenges by proposing a closed-loop automation and orchestration framework that ensures resilience against DoS attacks while maintaining service quality even under adverse conditions.

The proposed framework employs real-time monitoring, intrusion detection, SDN, and NFV to provide an integrated and automated response to network anomalies. A key aspect of this approach is the simultaneous management of control- and user-plane functions, enabling coordinated mitigation strategies to prevent service degradation during attack detection and response. Unlike traditional reactive approaches, this framework operates in a closed loop, where monitoring, detection, decision-making, and enforcement occur continuously in an automated cycle. This allows the system to dynamically instantiate additional VNFs to compensate for resource shortages, while enforcing security policies against malicious traffic. Consequently, mitigation actions are performed in parallel with service provisioning to ensure that critical communication is preserved.

The system architecture is centered on an intrusion detection system that continuously analyzes traffic in both control and user planes. When anomalous behavior is detected, the system generates alerts containing detailed information about the offending endpoints, such as IP addresses and subscriber identifiers. These alerts are forwarded to an SDN controller that applies appropriate enforcement rules by configuring network elements, such as firewalls or radio access nodes. Simultaneously, the orchestration layer dynamically allocates additional resources by instantiating VNFs, thereby alleviating congestion and maintaining service performance. This architectural approach enables a closed feedback loop in which detection and mitigation are tightly coupled, providing a high degree of automation and adaptability across segments of the network, including the radio access network, edge, and core, and potentially extending to satellite domains.

The experimental validation of the framework was conducted using a comprehensive testbed that integrated both commercial and open-source tools to emulate realistic 5G network conditions (see Figure 21). Traffic generation, attack simulation, and service provisioning are implemented using a combination of platforms that can reproduce user-plane and control-plane procedures, as well as VoIP services. Monitoring and intrusion detection are performed through deep packet inspection and real-time analytics, while orchestration mechanisms are implemented via programmable network control. This setup enables the evaluation of the framework across diverse, representative attack scenarios, including signaling storms, high-volume data floods, and application-layer disruptions affecting voice and video services.

Three main DoS scenarios are considered to assess the effectiveness of the proposed approach. In the first scenario, control-plane attacks arise from excessive signaling requests that overwhelm the network functions responsible for session management and access control (specifically, AMF). In the second scenario, user-plane attacks originate from malware-infected endpoints that generate high-volume traffic, saturate bandwidth, and overload the UPF data-forwarding functions. In the third scenario, application-layer attacks target SIP-based VoIP services, degrading communication quality and preventing the establishment of new sessions. Figure 22 shows the results of this type of attack, which degrades the QoS for voice traffic. It can be observed that this degradation is minimized once the attack is detected and mitigated. In all cases, the intrusion detection system successfully identifies anomalous behaviors and triggers enforcement mechanisms that block or limit malicious traffic based on the identity- or application-level characteristics. Simultaneously, the orchestration system compensates for the increased load by dynamically allocating the additional resources.

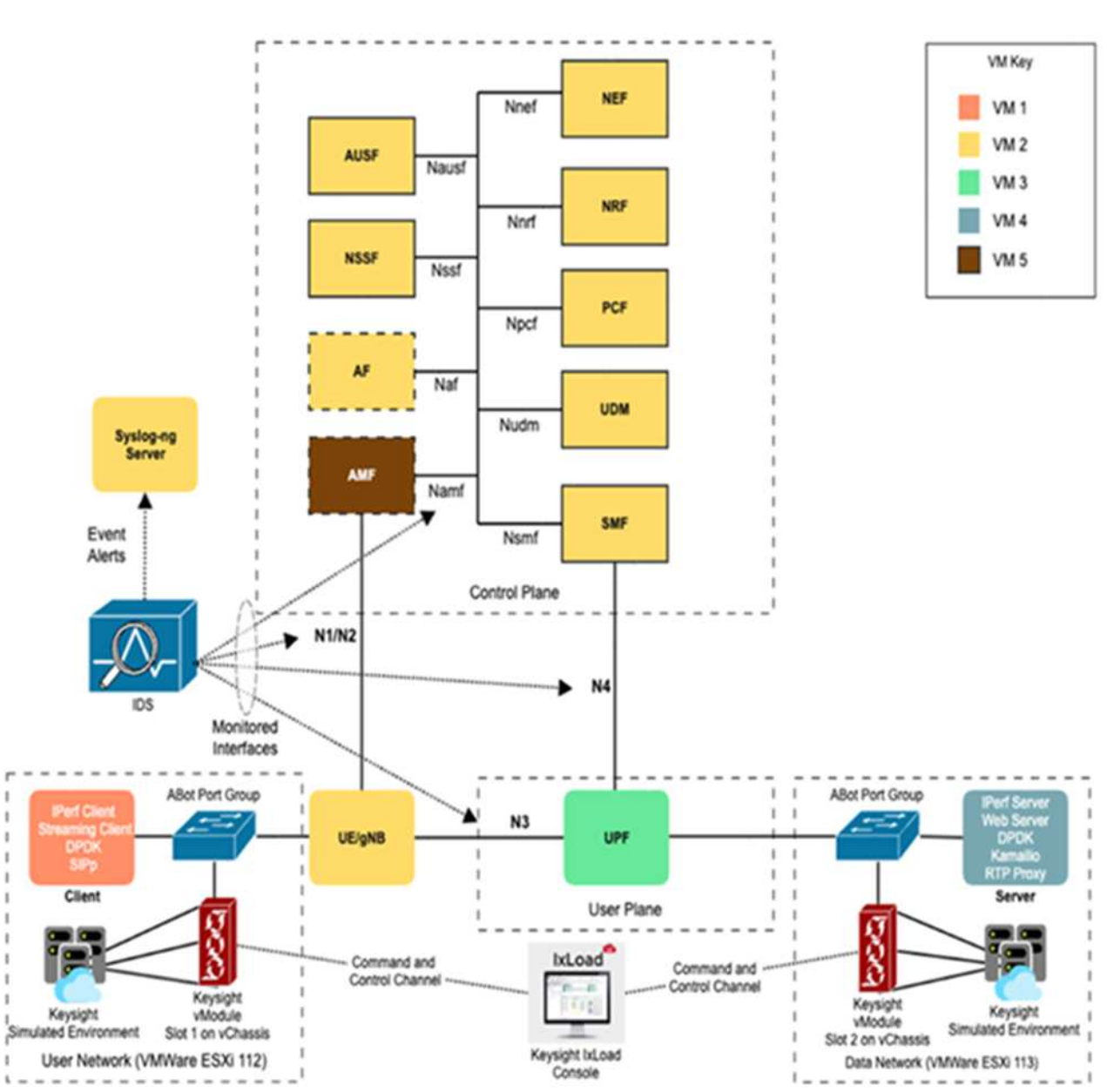

FIGURE 21. **Closed-loop automation and orchestration testbed.**

The results demonstrate that DoS attacks significantly affect network performance, as evidenced by the increased packet loss and reduced quality-of-experience metrics, such as the Mean Opinion Score (MOS) for voice traffic. However, once the closed-loop framework detects an attack and applies mitigation strategies, the system can effectively restore service quality. Packet loss decreases and quality metrics return to acceptable levels, indicating that the combined use of detection, enforcement, and dynamic orchestration successfully maintains service continuity. Importantly, this recovery occurs without interrupting the ongoing communication, highlighting the advantages of performing mitigation and resource allocation in parallel.

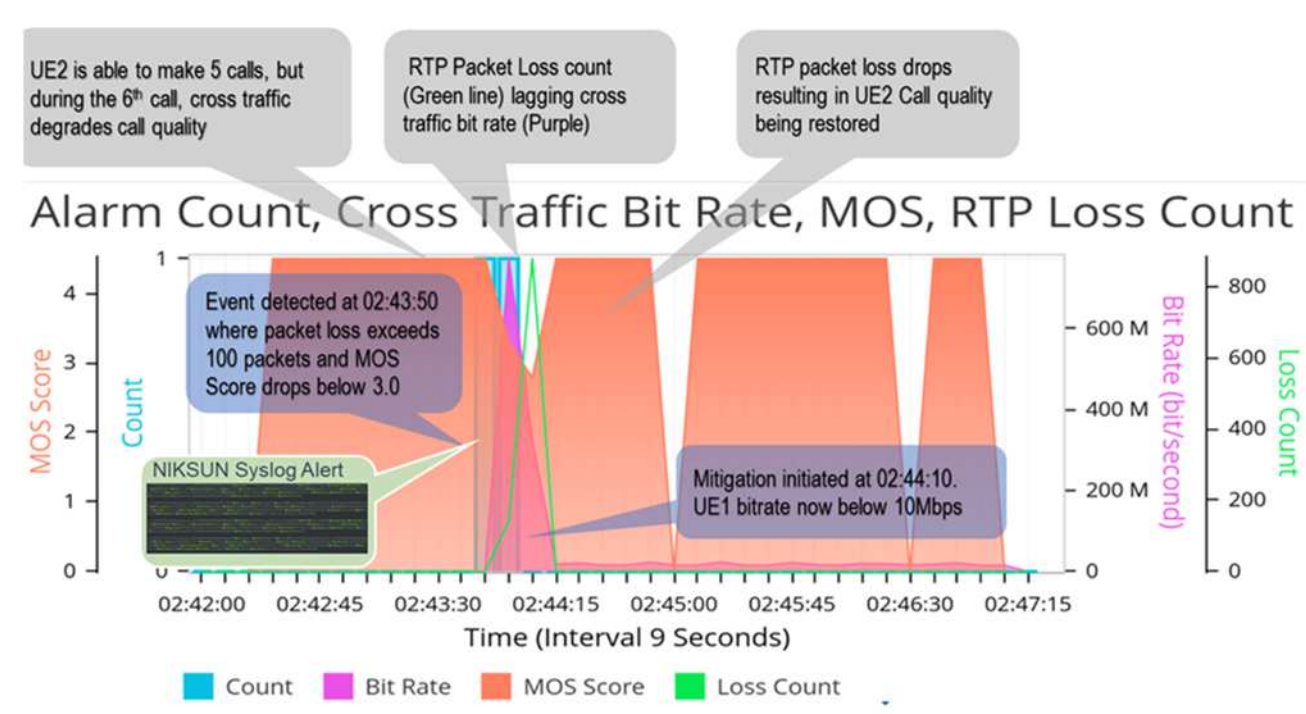

FIGURE 22. **Results of attacks and mitigation for SIP-based denial of service attacks.**

By ensuring that detection, mitigation, and orchestration are tightly integrated within a closed-loop system, this approach supports the vision of autonomous network management for next-generation integrated communication networks. The ability to maintain service continuity under attack conditions while efficiently utilizing network resources represents a significant step toward realizing a robust and secure communication infrastructure for future 6G networks.

## X. CLOUD-NATIVE ARTIFICIAL INTELLIGENCE-BASED SECURITY ARCHITECTURE

Advances in cloud computing and web development over the last 20 years have led to the emergence of many significant systems, technologies, and applications. For example, computing capabilities via containerization and Kubernetes, data processing architectures such as Apache Spark and Apache Beam, AI and ML techniques employing Deep Neural Networks (DNNs), and the most recent Generative AI (GenAI) and its companion, Agentic AI, have led to breakthroughs in many areas such as e-commerce, telemedicine, synthetic images, and video generation. The next step is to translate these innovations into practical cybersecurity solutions for future 6G T-NTN, enabling more resilient and automated security architectures.

Even with such great innovation, threats from globally connected worlds have increased. As such, modern cybersecurity architectures and techniques must evolve. To accomplish this objective, Zero Trust (ZT) security aims to mitigate these threats by implementing proven techniques to harden network infrastructure, as discussed in [85]. Here, the focus is on implementing cybersecurity from a cloud-native perspective.

Cloud-native design, where a definition can be taken from the Cloud Native Computing Foundation (CNCF) [86], generally refers to the use of microservices, containerization, orchestration, and more, to allow cloud resources to be provisioned in a scalable and agile manner, often via API calls. From a computing perspective, scalability and agility, which are key to 6G networks, can be achieved using Kubernetes and containers. Scalability and agility for data can take the form of massive databases, such as Google's BigQuery, data ingestion and pipelines via Apache Beam, etc. 6G T-NTN must follow the same approach to build autonomy, scalability, and agility into their networks.

Why is cloud-native so important? Future data centers are increasingly adopting cloud-native design principles to improve scalability, flexibility, and resource utilization. However, today's public cloud providers, including Google Cloud, Microsoft Azure, and Amazon Web Services (AWS), are already struggling to meet the rapidly growing demand for computing resources. This challenge is particularly evident for Generative AI (GenAI) workloads, where demand currently far exceeds the available supply, and this is only the beginning of the GenAI revolution. Demand will increase exponentially, along with the computing resources required to support it. However, this is only a part of the problem. Owing to environmental concerns and the power consumption of data centers, building new data centers is becoming increasingly challenging [87],[88]. Hence, the race to deploy data centers in space has begun [89]. The advent of cloud computing and cloud-native design over the past two decades has brought significant changes to applications and system development and has provided a strong foundation for 6G networks. While implementation is always the most challenging part of security design, starting with a cloud-native basis and building from there provides a better roadmap, especially when one considers that the future 6G TNs and NTNs will be integrated with public cloud providers.

Future cybersecurity architectures must integrate these technologies to secure networks, particularly as they evolve to address advances in computing and networking infrastructure, and emerging AI/ML capabilities. The core components of a 6G cybersecurity architecture include a robust data processing architecture capable of handling large volumes of streaming and batch data, an AI/ML architecture that integrates advanced algorithms, such as cyber-based transformer models [90], and a GenAI-based system for autonomy and orchestration.

As an example of a data-processing architecture for secure 6G networks, Figure 23 presents a conceptual architecture for autonomous security processing that integrates multiple data sources. The figure illustrates both batch and stream processing with multiple instances for each. Each stream is processed using a pipeline that cleans and distributes the data. For simplicity, only a data distribution network is illustrated; however, significant database processing must occur before the data are sent to an endpoint. Database choices depend on the specific type of threat being monitored and on read/write latency. For fixed schemas, an SQL database may be sufficient; however, for high-rate data, such as IoT streaming, a NoSQL database with a flexible schema may be the choice. Google's Bigtable, on which Apache HBase is modeled, provides a very high throughput and has a flexible schema. There are many choices depending on the type of threat being monitored. Hence, many trade-offs must be analyzed for a particular system.

Similarly, Figure 24 illustrates the cloud-native AI/ML cybersecurity processing architecture. Here, an agentic orchestrator controls and manages the security architecture, which consists of task agents housed in the cloud infrastructure itself, while also connecting with external agents that may be part of other networks that provide services. Model Context Protocol (MCP) is used to connect external agents and task agents to external tools, such as databases and other AI/ML models.

From these two figures, it is clear that there are many new opportunities related to cloud-native design for both data-processing and AI/ML security architectures.

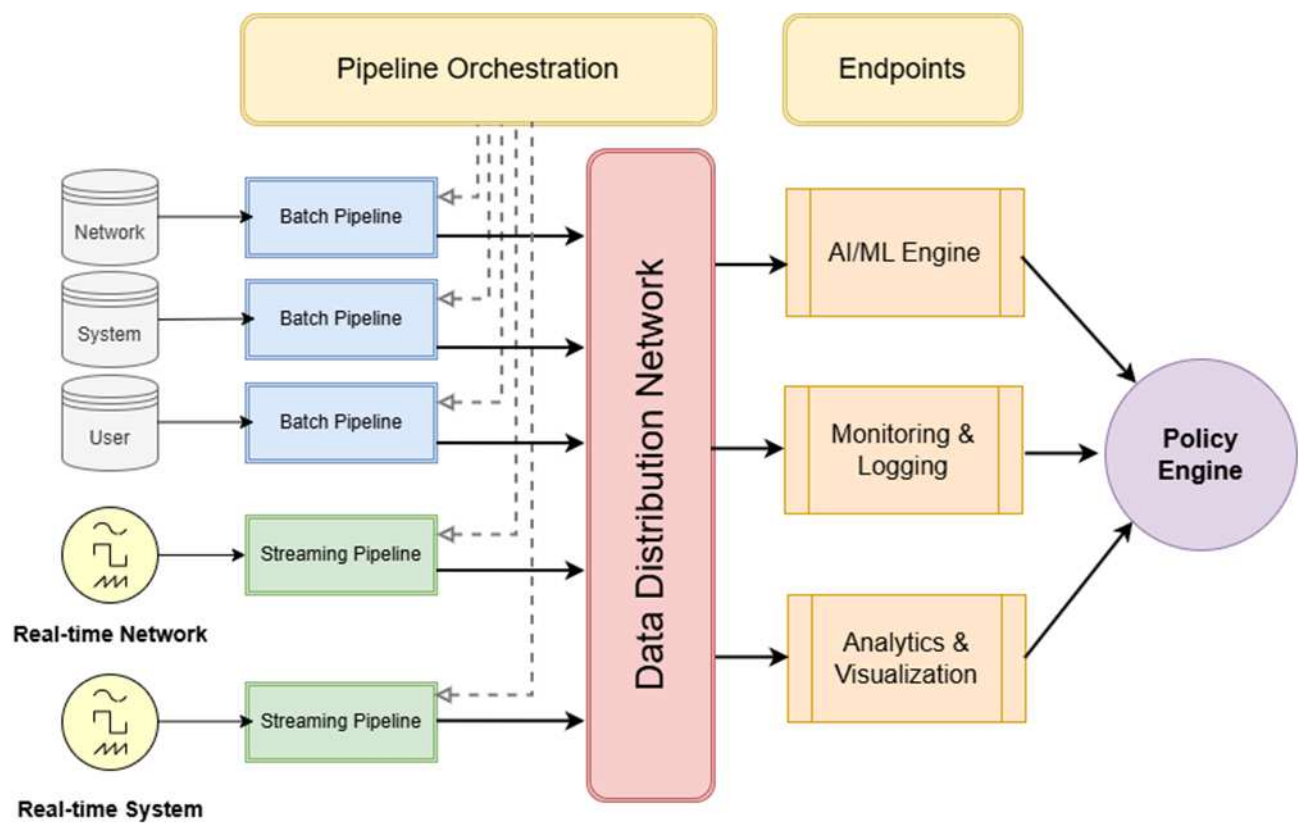

**FIGURE 23.** **Cloud-native data processing architecture for cybersecurity orchestration and automation.**

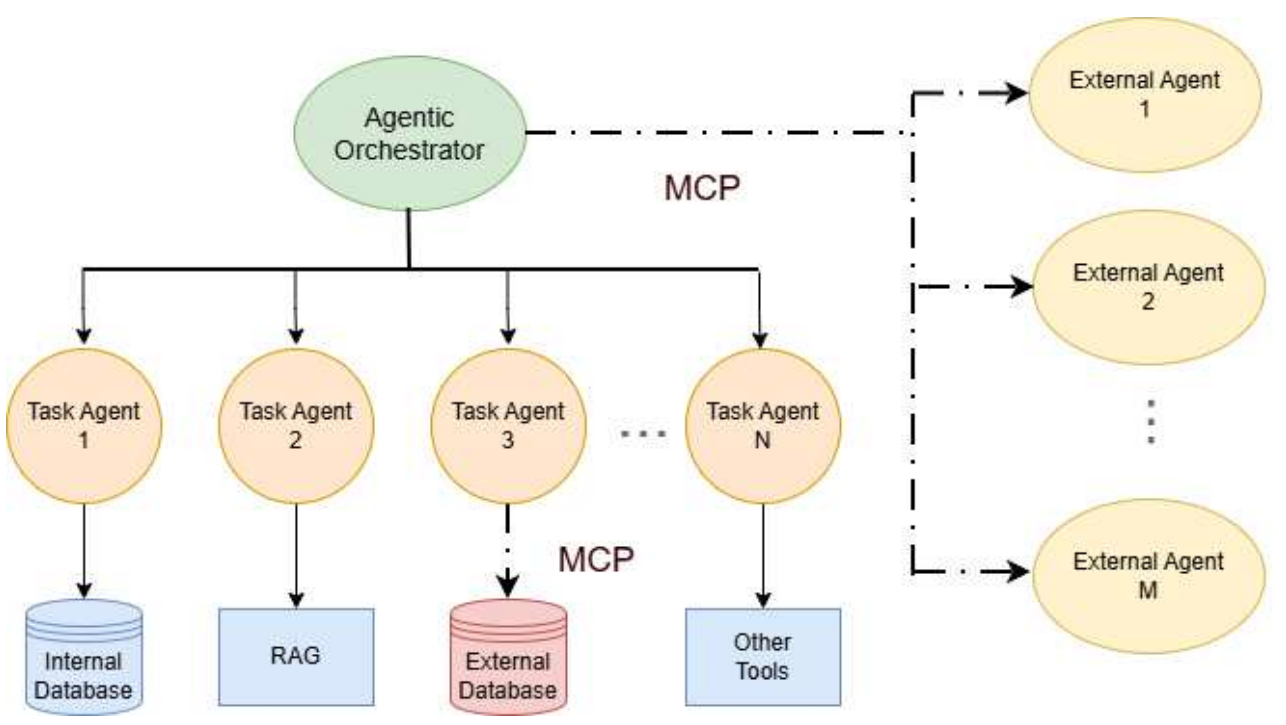

**FIGURE 24.** **Cloud-native AI/ML processing architecture for cybersecurity orchestration and automation.**

## XI. QUANTUM SECURITY ARCHITECTURE AND INTEROPERABILITY

The evolution toward a fully federated T-NTN architecture must consider future security challenges, including the significant threats posed by quantum computers, which can render today's encryption obsolete. Post-Quantum Cryptography (PQC) offers a family of classical cryptographic algorithms designed to remain secure against quantum attacks, while Quantum Key Distribution (QKD) exploits the principles of quantum mechanics to enable cryptographic key exchange with security guaranteed by the laws of physics [91]. Integrating such an unbreakable global-scale QKD security layer above the PQC layer is paramount for protecting sensitive diplomatic, military, and financial communications. QKD is especially vital against "harvest now, decrypt later" strategies, in which communications stored for years or decades could be decrypted if future techniques, such as quantum computers, break deployed PQC. QKD will also become critical for in-space communications as future ODCs and other sensitive space assets are deployed in orbit and integrated into federated network infrastructures.

Satellite quantum communication transmits quantum information through single photons. The spectacular success of the Micius satellite, which demonstrated satellite-based quantum entanglement distribution between ground stations separated by more than 1,200 km in 2017 [92], and the subsequent Jinan-1 microsatellite mission in 2024 [93] have demonstrated the feasibility of satellite-based quantum communications, thus leading many countries to announce their own quantum satellite missions [94],[95]. Active research is currently focused on achieving global coverage (~20,000 km) [94],[95]. Several directions are under debate, including the use of high-orbit satellites [96] with lower data rates or low-orbit satellite networks [95],[97],[98] that offer higher rates but require more resources and sophisticated technologies. Within low-orbit networks, two primary technological directions are emerging: satellite-based quantum repeaters using sophisticated quantum memories [94],[97], and satellite-chain reflection technology [95],[98], which can enable long-distance communication via reflection, bypassing the need for quantum memories or repeaters.

Broadly, quantum communication is very similar to optical laser links, although it operates at a low power at the single-photon level. Beyond high data rates, classical optical communication inherently offers high security because its narrow beams are much more difficult to intercept than radio-wave transmissions, which have a significantly larger receiving area. To ensure interoperability, quantum and classical optical communication can be integrated using two main methods. In the "coexistence" approach, quantum and classical signals travel through the same optical system but are separated in wavelength, time, or space using specialized filters [99]. In simultaneous classical and quantum communication approaches, quantum information is directly embedded within a classical signal [100]. Real-world demonstrations of the coexistence approach using the Jinan-1 satellite have shown that such integration can significantly reduce the hardware size and improve deployment speed [94].

The European Telecommunications Standards Institute (ETSI) Industry Specification Group on Quantum Key Distribution (ISG-QKD) leads QKD standardization efforts in Europe and spearheads the development of multi-vendor interface standards to ensure seamless interoperability between satellite nodes and the ground segment. These efforts are pivotal to the architectural integration of sovereign quantum networks, providing the standardized key management framework required to support cross-border initiatives such as EuroQCI and the FranceQCI pilot.

## XII. CONCLUSIONS

At the end of this work, which addresses several key issues for future integrated T-NTN systems, a coherent perspective emerges around four tightly coupled dimensions.

From an **architectural standpoint, the evolution toward 6G T-NTNs converges on federated, cloud-native**, and service-based designs in which terrestrial and satellite segments operate as a unified yet distributed system. The federation significantly improves routing efficiency and resilience (achieving up to 23% reduction in average delay, complete elimination of call blocking, and nearly 47% reduction in worst-case latency) while enabling advanced orchestration, service chaining, and cross-domain resource management. This shift is essential for supporting global coverage, seamless service continuity, and heterogeneous deployments across terrestrial, aerial, and space domains.

Within this framework, **O-RAN** plays a central role. Its disaggregated and programmable architecture enables edge intelligence through the hierarchical RIC ecosystem (Non-RT and Near-RT), enhances flexibility via dynamic placement of CU/DU/RIC functions and applications across TN and NTN segments, improves energy efficiency by optimizing the allocation of computing and communication resources, and supports network slicing for diverse QoS requirements.

In parallel, **a multi-operator federation** reshapes the security paradigm. Traditional perimeter-based security models are being replaced by the zero-trust approach, in which all architectural entities (satellites, gateways, edge nodes, and administrative domains) are continuously authenticated, authorized, and verified. This requires strong identity management, end-to-end protection, secure cross-domain policy exchange, and AI-driven closed-loop security mechanisms that are capable of real-time detection and mitigation, ensuring service continuity even under attack.

**Quantum threats** are of critical concern. While PQC provides near-term protection, it may be insufficient for long-lived assets such as satellites. In this context, QKD has emerged as a complementary solution, enabling secure key exchange based on physical principles.

Taken together, these architectural, operational, and security challenges highlight the need for a coordinated effort to guide the evolution of future integrated T-NTN systems. In this context, the IEEE INGR Satellite Working Group can play an important role by fostering collaboration among academia, industry, and standardization bodies, and by aligning its activities with the IMT-2030 Vision to help shape the next generation of T-NTN systems.

## ACKNOWLEDGMENTS

The authors would like to acknowledge the support provided to the IEEE INGR Satellite WG and the INGR MHz-to-THz Networks WG by IEEE Future Networks. This publication is supported by the 5G-HUB project (HORIZON-EUSPA-2023-SPACE), funded within the EU Horizon Europe research and innovation program (grant agreement No. 101180143). This work was also partly supported by 5G-STARDUST, 6G-NTN, UNITY-6G, and NexaSphere Projects from the Smart Networks and Services Joint Undertaking (SNS JU) under the European Union's Horizon Europe Research and Innovation Program under Grants 101096573, 101096479, 101192650, and 101192912, and in part by the Swiss State Secretariat for Education, Research, and Innovation (SERI) and Xjenza Malta's support of the SMARTEN6G project. Finally, this work was supported in part by IIT Palakkad Technology IHub Foundation Doctoral Fellowship (IPTIF) under Grant IPTIF/HRD/DF/032/SEP46, in part by the Department of Science and Technology–Science and Engineering Research Board (IPTIF DST-SERB) under Grant CGR/2021/009286, and in part by the Scheme for Promotion of Academic and Research Collaboration (SPARC) Project under Grant P3701.

The views expressed are those of the authors, and do not necessarily represent the projects mentioned.